\documentclass[10pt, prd, aps, amssymb, amsmath, tightenlines, noshowkeys, 
twocolumn]{revtex4}

\usepackage{graphicx}
\usepackage[T1]{fontenc}
\usepackage[latin1]{inputenc}
\usepackage[english]{babel}
\usepackage{amsmath}
\usepackage{amsfonts}
\usepackage{amssymb}
\usepackage{amsthm}
\usepackage{dsfont}
\usepackage{color}
\usepackage{xcolor}
\usepackage{float}
\usepackage{afterpage}
\usepackage{slashed}
\usepackage[labelformat=parens]{subfig}
\usepackage{bbm}
\usepackage[colorlinks=true, linkcolor=blue, urlcolor=blue, citecolor=blue,
            anchorcolor=blue]{hyperref}
\usepackage[justification=raggedright, font=footnotesize]{caption}

\def\cP{\mathcal P}
\def\cT{\mathcal T}
\def\cPT{\mathcal{PT}}
\def\cH{\mathcal{H}}

\date{\today}

\begin{document}

\title{Fermion and meson mass generation in non-Hermitian Nambu--Jona-Lasinio models}

\author{Alexander Felski}\email{felski@thphys.uni-heidelberg.de}
\author{S.~P. Klevansky}\email{spk@physik.uni-heidelberg.de}
\affiliation{Institut f\"{u}r Theoretische Physik, Universit\"{a}t Heidelberg,
Philosophenweg 12, 69120 Heidelberg, Germany}

\begin{abstract}
We investigate the effects of non-Hermiticity on interacting fermionic systems. We do this by including non-Hermitian bilinear terms into the 3+1 dimensional Nambu--Jona-Lasinio 
(NJL) model. Two possible bilinear modifications give rise to $\cPT$ symmetric theories; this happens when the standard NJL model is extended either by a pseudovector 
background field $ig \bar\psi\gamma_5 B_\mu \gamma^\mu \psi$ or by an antisymmetric-tensor background field $g \bar\psi F_{\mu\nu}\gamma^\mu \gamma^\nu \psi$. The three 
remaining bilinears are {\it anti}-$\cPT$-symmetric in nature, $ig \bar\psi B_\mu \gamma^\mu \psi, ig\bar\psi \gamma_5 \psi$ and $ig\bar\psi \mathbbm{1}\psi$, so that 
the Hamiltonian then has no overall symmetry. The pseudovector $ig \bar\psi\gamma_5 B_\mu \gamma^\mu \psi$ and the vector $ig \bar\psi B_\mu \gamma^\mu \psi$ combinations, 
are, in addition, chirally symmetric. Thus, within this framework we are able to examine the effects that the various combinations of non-Hermiticity, $\cPT$ symmetry, chiral symmetry
and the two-body interactions of the NJL model have on the existence and dynamical generation of a real effective fermion mass (a feature which is absent in the corresponding 
modified massless free Dirac models) as well as on the masses of the composite particles, the pseudoscalar and scalar mesonic modes ($\pi$ and $\sigma$ mesons). Our findings
demonstrate that $\cPT$ symmetry is not necessary for real fermion mass solutions to exist, rather the two-body interactions of the NJL model supersede the non-Hermitian bilinear 
effects. The effects of chiral symmetry are evident most clearly in the meson modes, the pseudoscalar of which will always be Goldstone in nature if the system is chirally symmetric. 
Second solutions of the mesonic equations  are also discussed.
\end{abstract}

\maketitle

\section{Introduction}
\label{s1}

The occurrence of real eigenvalue spectra in non-Hermitian 
systems that are symmetric under combined 
parity reflection $\cP$ and time reversal $\cT$ has, since its first
demonstration by Bender and Boettcher in 1998 \cite{bb}, inspired a wide 
variety of theoretical and experimental studies in $\cPT$-symmetric physics. 
While initially focusing on classical and semi-classical systems these 
studies were soon extended to include bosonic field theories as well 
\cite{bhkss}.
For fermionic systems, Jones-Smith et al.~\cite{jsm} 
brought attention to the property of odd time-reversal symmetry 
($\cT^2=-\mathbbm{1}$) in $3+1$ dimensions and its importance in 
the context of $\cPT$ symmetry, 
in contrast to the even time-reversal symmetry of bosonic systems.

In further studies involving fermions, unexpected behavior has emerged:
In \cite{bkb} the free Dirac Lagrangian was modified by the addition
of non-Hermitian, but $\cPT$-symmetric, bilinears of the fermionic 
field $\psi$ and its conjugate $\bar\psi$.
It was found that an unbroken $\cPT$ symmetry 
phase, that is, a regime with a real spectrum, does not simply exist: 
without a finite bare mass these modified Dirac fermions generally have
\textit{complex} physical masses. 
However, including  higher-order interactions can override this breakdown 
of an unbroken $\cPT$ symmetry regime. 
We demonstrated in \cite{fbk} that a 
region of real mass solutions exists, when the bilinear term that  
includes the $\cPT$-symmetric pseudovector background field 
$ig \bar\psi \gamma_5 B_\mu \gamma^\mu \psi$, is incorporated into the 
$3+1$ dimensional Nambu--Jona-Lasinio (NJL) model, which contains
two-body interactions as well. 
The gap equation of this modified system with Hamiltonian density
\begin{equation}
\label{s1e1}
\cH=\bar\psi (-i \gamma^k \partial_k + m_0 +ig \gamma_5 B_{\mu}\gamma^{\mu} )
\psi - G [ (\bar\psi \psi)^2 + (\bar\psi i \gamma_5 \vec{\tau} \psi)^2 ]
\end{equation}
has real physical mass solutions even in the chiral limit of vanishing bare 
mass $m_0$, provided the coupling constant $g$ of the non-Hermitian bilinear 
does not exceed a critical value $g_{crit}$. 
Moreover, when compared to the standard NJL model ($g=0$), increased fermion 
masses could be generated dynamically, mimicking the effect of a small 
bare mass $m_0$ without breaking chiral symmetry.

Similar results have been demonstrated in \cite{ams,ms,ccr} for models with 
Yukawa-type interactions that couple different fields,
and can at least in principle be 
obtained from four-point interactions through partial bosonization.

The existence of an unbroken $\cPT$ symmetry regime in the modified 
NJL model (\ref{s1e1}) without a bare mass $m_0$ is remarkable, 
given that in the underlying modified Dirac theory ($G=0$) this symmetry is
realized in the broken regime exclusively.
It raises the question as to whether the addition of a non-Hermitian bilinear 
has to uphold $\cPT$ symmetry at all.
We thus ask what specific role the combination of chiral symmetry, $\cPT$
symmetry and higher-order interactions play in the
generation of real masses within fermionic models.
The first aim of this study is to expand the analysis of the 
modified NJL model (\ref{s1e1}) to cover all possible non-Hermitian bilinear extensions,
including but not limited to the only other $\cPT$-symmetric case of an antisymmetric-tensor 
background field $g \bar\psi F_{\mu\nu} \gamma^\mu \gamma^\nu \psi$ 
\cite{bkb}.
The analysis of these modifications to the NJL model furthermore covers
cases in which chiral symmetry is preserved as well as cases in which it is 
broken explicitly.

Our second aim in this paper goes beyond the discussion of
fermionic mass generation to study the 
effects on the mesonic composite states that can be generated within 
the context of these models.
For those systems in which we find dynamically generated real 
fermion masses, we investigate the effect that the non-Hermitian bilinear terms
have on the 
mass of the scalar and pseudoscalar bound states, i.e. these are the 
$\sigma$ and $\pi$ mesons in the context of quantum chromodynamics (QCD). 
 
This paper is structured as follows: 
In Sec.~\ref{s2} the modified NJL model is introduced and all
possible bilinear additions leading to a non-Hermitian model are identified.
Their behavior under $\cPT$ symmetry and chiral 
symmetry is presented. 
In Sec.~\ref{s3} we first recapitulate the algebraic form of the gap equation 
and its solution for the system previously studied in \cite{fbk}. 
We  then proceed to analyze the structure of the gap equations and the 
existence of real fermion mass solutions for the other non-Hermitian 
extensions of the NJL model.
In Sec.~\ref{s4} we investigate the corresponding mass of the scalar and
pseudoscalar bound states for models in which a dynamical fermion mass 
generation could be identified. 
We conclude in Sec.~\ref{s5}.

\section{The modified NJL model}
\label{s2}

In its two-flavor version ($N_f=2$), the standard NJL model \cite{njl} is
given by the Hamiltonian density
$$
\mathcal{H}_{\textrm{NJL}} = \bar\psi (-i \gamma^k \partial_k + m_0) \psi 
- G [ (\bar\psi \psi)^2 + (\bar\psi i \gamma_5 \vec{\tau} \psi)^2 ],
$$
in terms of the  Dirac matrices $\gamma$ with $k=1..3$, the isospin SU(2) matrices 
$\vec{\tau}$, a coupling strength $G$, and the bare mass $m_0$. 
While the mass term $m_0$ breaks chiral symmetry explicitly, the specific combination of four-point interaction terms preserves it. 
In the limit of vanishing bare mass the model can thus be used to study the 
spontaneous breaking of chiral symmetry occurring through a mechanism that
parallels pair condensation in the Bardeen-Cooper-Schrieffer (BCS) theory of 
superconductivity \cite{bcs}.
As such it has been developed into an effective field theory of QCD in the 
chiral limit.
In this context, including a small bare mass $m_0$ leads to a model with an \emph{approximate} spontaneously broken chiral symmetry that has a small current 
quark mass and a light pseudo-Goldstone  boson, identified as the pion.
The numerical analysis throughout this paper will be based on 
established quantities within the QCD interpretation, using a four-momentum Euclidean cutoff scale of $\Lambda=1015$~MeV for the purpose of regularization and a 
coupling strength satisfying $G \Lambda^2= 3.93$, cf.~\cite{spk}.

In continuation of our previous study \cite{fbk}, we introduce the 
bilinear extension of the NJL model
\begin{equation}
\label{s2e2}
\cH= \bar\psi (-i \gamma^k \partial_k + m_0+g \Gamma ) \psi 
- G [ (\bar\psi \psi)^2 + (\bar\psi i \gamma_5 \vec{\tau} \psi)^2 ]  ,
\end{equation}
where $\Gamma$ generally denotes a complex $4\times4$ matrix and $g$ is the 
associated real coupling constant. 
In the following, we consider the complete set of $4\times4$ matrices generated by the Dirac 
matrices and identify all those that are non-Hermitian, irrespective of whether
they are $\cPT$-symmetric or not. 
The matrix $\Gamma$ in (\ref{s2e2}) will then take on the corresponding role in the 
subsequent analysis. 
In addition, we consider the behavior of the non-Hermitian bilinears under chiral symmetry transformations, and thus infer the 
symmetries present in the modified NJL models given by (\ref{s2e2}).

Any bilinear of the fermionic field $\psi$ and its conjugate $\bar\psi$ is of 
the form $\bar\psi \Gamma \psi$. 
As such, they can be written as real superpositions based on the 
following $32$ matrices:
\begin{equation}
\label{s2e3}
\begin{alignedat}{5}
&\mathbbm{1},&& \quad 
\gamma_5,&& \quad
B_\mu\gamma^\mu,&& \quad
\gamma_5 B_\mu\gamma^\mu,&& \quad 
F_{\mu\nu}\gamma^\mu \gamma^\nu,\\
&i\mathbbm{1},&& \quad 
i\gamma_5,&& \quad
iB_\mu\gamma^\mu,&& \quad
i\gamma_5 B_\mu\gamma^\mu,&& \quad 
iF_{\mu\nu}\gamma^\mu \gamma^\nu,
\end{alignedat}
\end{equation}
where $\mu\leq\nu$ denote spin indices, $B_\mu$ are real vector elements, and $F_{\mu\nu}$ are 
real elements of an antisymmetric matrix.
The corresponding bilinears  $\bar\psi \Gamma\psi$ formed by these behave  as scalars, pseudoscalars, 
vectors, pseudovectors and antisymmetric $2$nd-rank tensors  respectively under Lorentz 
transformations (from left to right).
Including an imaginary unit into any bilinear will 
change it from being symmetric to being antisymmetric under Hermitian 
conjugation, and vice versa. 
Thus we identify half of the bilinear terms associated with the
matrices in (\ref{s2e3}) to be anti-Hermitian, namely those for which 
$\Gamma$ takes one of the forms
\begin{equation}
\label{s2e4}
i\mathbbm{1}, \quad
\gamma_5, \quad
i B_\mu\gamma^\mu, \quad
i\gamma_5 B_\mu\gamma^\mu, \quad
F_{\mu\nu}\gamma^\mu \gamma^\nu .
\end{equation}
These terms result in the non-Hermitian extensions of the NJL 
model considered throughout this paper.

To determine the behavior of $\bar\psi \Gamma \psi$, with $\Gamma$ being one 
of the terms in (\ref{s2e4}), under combined parity reflection $\cP$ and time 
reversal $\cT$, recall the usual definition of these transformations in $3+1$ dimensions \cite{bd}:
\begin{alignat}{2}
\label{s2e5}
\cP:&\, \psi(t, \mathbf{x}) \to \cP \psi(t, \mathbf{x}) \cP^{-1} &&= 
\gamma^0 \psi(t, -\mathbf{x}),\\
\label{s2e6}
\cT:&\, \psi(t, \mathbf{x}) \to \cT \psi(t, \mathbf{x}) \cT^{-1} &&= 
i \gamma^1 \gamma^3 \psi^*(-t, \mathbf{x}).
\end{alignat} 
We find that only two types of non-Hermitian bilinears are 
$\cPT$-symmetric; 
specifically, $[\cPT, \Gamma]=0$ for
\begin{align}
\label{s2e7}
\Gamma_{PT_1} &= i\gamma_5 B_\mu\gamma^\mu , \\
\label{s2e8}
\Gamma_{PT_2} &= F_{\mu\nu}\gamma^\mu \gamma^\nu.
\end{align}
The remaining non-Hermitian terms are anti-$\cPT$-symmetric, 
$\{\cPT, \Gamma\}=0$, and will in the following be referred to as 
\begin{align}
\label{s2e9}
\Gamma_{aPT_1} &= i B_\mu\gamma^\mu , \\
\label{s2e10}
\Gamma_{aPT_2} &= \gamma_5 , \\
\label{s2e11}
\Gamma_{aPT_3} &= i\mathbbm{1}.
\end{align}
Notice that in these anti-$\cPT$-symmetric cases the Hamiltonian density 
(\ref{s2e2}) has no such overall symmetry. The system as a whole is thus non-Hermitian and 
non-$\cPT$-symmetric.

Moreover, only two of the terms in (\ref{s2e4}) anticommute 
with $\gamma_5$, leading to an axial flavor symmetry and thus an overall 
chiral symmetry of the Hamiltonian density (\ref{s2e2}) in the limit of vanishing bare 
mass: $\Gamma_{PT_1}$ and $\Gamma_{aPT_1}$. 
In contrast, even without a bare mass $m_0$, chiral symmetry is  broken 
explicitly in the theories based on including
the terms $\Gamma_{PT_2}$, $\Gamma_{aPT_2}$, and $\Gamma_{aPT_3}$.

Thus, we have identified five possible non-Hermitian bilinear extensions of 
the NJL model:
One that is $\cPT$-symmetric and chirally symmetric ($\Gamma_{PT_1}$), one that 
is $\cPT$-symmetric but breaks chiral symmetry explicitly ($\Gamma_{PT_2}$), 
one that is anti-$\cPT$-symmetric but chirally symmetric ($\Gamma_{aPT_1}$),
and two which break both $\cPT$ and chiral symmetry 
($\Gamma_{aPT_2}$ and $\Gamma_{aPT_3}$).

\section{Gap equation and fermion masses}
\label{s3}

A self-consistent approximation to the effective fermion mass of the standard
NJL model is expressed by the gap equation, which can be 
obtained following Feynman-Dyson perturbation theory, 
cf.~\cite{spk,fw}.
We established in \cite{fbk} that the gap equation of an extended NJL model
such as (\ref{s2e2}), which includes additional bilinear terms, has the same 
general structure as that of the standard NJL model: 
\begin{equation}
\label{s3e1}
m = m_0 + \frac{2 i G N_c N_f}{(2\pi)^4} \, I,
\end{equation}
where $m$ denotes the effective fermion mass, $N_c=3$ the number of colors, 
$N_f=2$ is the number of flavors, and $I$ is the momentum integral over the spinor trace of the fermion propagator:
\begin{equation}
\label{s3e2}
I=  \int \!\!d^4 p \,\, {\rm tr}[S(p)].
\end{equation}
Here, the full fermion propagator $S(p)$ has the same form as the propagator
$S^{0}(p)$ of the corresponding modified Dirac theory (obtained at $G=0$) and can be found by 
replacing the bare mass $m_0$ with the effective mass $m$:
\begin{equation}
\label{s3e3}
S(p) = (\slashed p - m-  g \Gamma)^{-1} .
\end{equation}
Analyzing the effective fermion mass of the non-Hermitian NJL models introduced in the previous section thus relies on the 
evaluation of the spinor trace and the momentum integration in (\ref{s3e2})
for these models. 

In this section we derive the algebraic gap equation for the five non-Hermitian NJL models and study the existence and behavior of real 
fermion mass solutions. 
Section~\ref{s3a} summarizes the results for the $\cPT$-symmetric 
model based on $\Gamma_{PT_1}$, which were previously obtained in \cite{fbk}.
In Sec.~\ref{s3b} the second $\cPT$-symmetric case (based on $\Gamma_{PT_2}$) is analyzed, and 
Secs.~\ref{s3c}-\hyperref[s3e]{E} discuss the three remaining non-$\cPT$-symmetric models.

\subsection{$\Gamma_{PT_1}=i\gamma_5 B_\mu \gamma^\mu$}
\label{s3a}

In the case of the $\cPT$-symmetric model based on including the pseudovector background field $\Gamma_{PT_1}=i\gamma_5 B_\mu\gamma^\mu$, as studied in \cite{fbk}, 
the full fermion propagator (\ref{s3e3}) was rationalized,
\begin{widetext}
\begin{equation}
\label{s3e3-2}
S(p) =
\frac{ 
(\slashed p + m + i\gamma_5 B_\mu\gamma^\mu)
\bigl[(p^2-m^2-g^2B^2) +2igm \gamma_5 B_\nu \gamma^\nu +2ig B_\nu p^\nu \gamma_5\bigr]
}{(p^2 - m^2 - g^2 B^2)^2 - 4 g^2 m^2 B^2
+ 4 g^2 (B_\mu p^\mu)^2},
\end{equation}
\end{widetext}
in order to evaluate the spinor trace
\begin{equation*}
\label{s3e4}
\mathrm{tr}[S(p)] = \frac {4 m (p^2 - m^2 + g^2 B^2)}
{(p^2 - m^2 - g^2 B^2)^2 - 4 g^2 m^2 B^2
+ 4 g^2 (B_\mu p^\mu)^2} .
\end{equation*}
The introduction of a Euclidean four-momentum cutoff $\Lambda$ for the purpose 
of regularization then allows the momentum integration in (\ref{s3e2}).
To that end we change to Euclidean coordinates with $p_0=ip_4$ and 
$B_0=iB_4$, such that $p^2=-p_E^2$, $B^2=-B_E^2$, and $B_\mu p^\mu = -B_E\cdot 
p_E$. 
In the chiral limit of vanishing bare mass $m_0$, the gap equation 
(\ref{s3e1}) then takes on the algebraic form
\begin{widetext}
\begin{align}
\label{s3e5}
\begin{split}
\frac{2 \pi^2}{\tilde{G} N_c N_f} = \frac{1}{4 \tilde{g}^2}
\bigg\{&
\sqrt{(1 + \tilde{m}^2 - \tilde{g}^2)^2 +4 \tilde{g}^2 \tilde{m}^2} 
(1 + \tilde{m}^2 + 7 \tilde{g}^2) - (\tilde{m}^2 + \tilde{g}^2)(2 + \tilde{m}^2 
+ 7 \tilde{g}^2) - 1\\
&
+4 \tilde{g}^2 (2 \tilde{g}^2 - \tilde{m}^2) \ln\Big[ \frac{1}{2 \tilde{m}^2} 
\Big( \sqrt{(1 + \tilde{m}^2 - \tilde{g}^2)^2 + 4 \tilde{g}^2 \tilde{m}^2} 
+ 1 + \tilde{m}^2 - \tilde{g}^2 \Big) \Big] \bigg\}
\end{split}
\end{align}
\end{widetext}
in terms of the rescaled 
parameters $\tilde{G}= G \Lambda^2$, 
$\tilde{m}=m \Lambda^{-1}$, and $\tilde{g}=g \lvert B_E\rvert \Lambda^{-1}$,
which is proportional to the amplitude of the pseudovector background field.

\begin{figure}
\centering
\includegraphics[width=0.48\textwidth]
{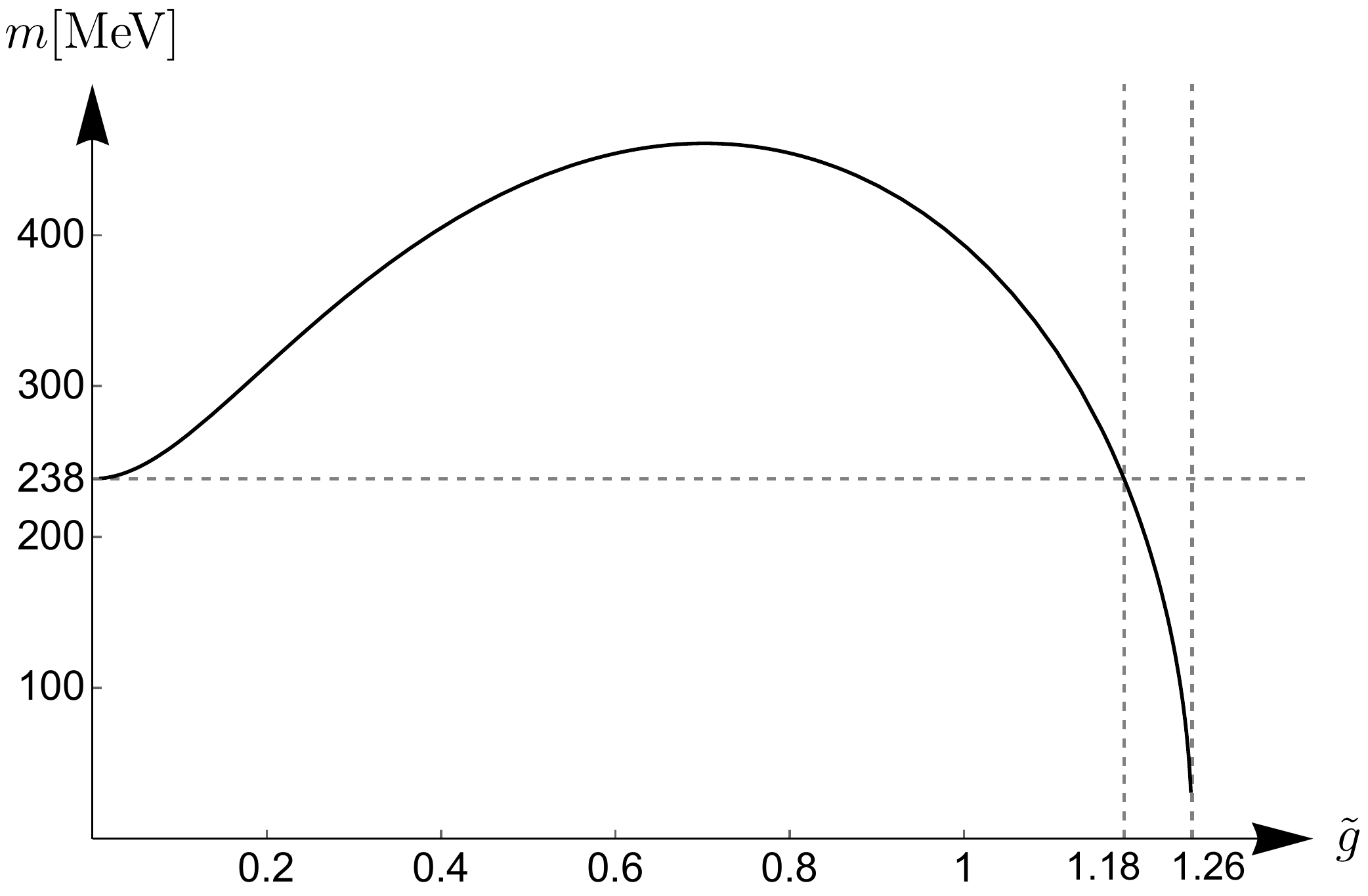}
\caption{ 
\label{f1}
Behavior of the effective fermion mass $m$ in the modified NJL model based on $\Gamma_{PT_1}$ 
as a function of the scaled coupling constant 
$\tilde{g}=g \lvert B_E\rvert \Lambda^{-1}$.}
\end{figure}

In the context of QCD, this equation has real fermion mass solutions $m$ 
as long as $\tilde{g}<\tilde{g}_{crit}\approx 1.261$. 
The behavior of the solution in this regime is shown in Fig.~\ref{f1}.
The largest mass of $m \approx 460.870$~MeV is found at a coupling value of 
$\tilde{g}\approx0.702$.
Notice that masses larger than the $m_{NJL}\approx 238.487$~MeV of the 
standard NJL model are dynamically generated for all coupling 
values below $\tilde{g}_{dyn}\approx 1.183$. 
A current quark mass in the range of the bare up quark mass, $m_u=(1.7-3.3)$~MeV,
is already generated at small coupling values 
$\tilde{g}\approx(0.025-0.034)$ 
and in the range of the bare down quark mass,
$m_d=(4.1-5.8)$~MeV, for $\tilde{g}\approx(0.038-0.046)$.

Altogether, the $\cPT$ and chirally symmetric modified NJL model based on 
$\Gamma_{PT_1}$ admits a finite region of unbroken $\cPT$ symmetry, even for 
vanishing bare mass $m_0$.
Real fermion mass solutions are obtained for coupling constant values
up to a critical value $\tilde{g}_{crit}$. 
Fermion mass is dynamically generated for 
$\tilde{g}<\tilde{g}_{dyn}<\tilde{g}_{crit}$; 
larger coupling values lead to an effective mass loss.

\subsection{$\Gamma_{PT_2}=F_{\mu\nu}\gamma^\mu \gamma^\nu$}
\label{s3b}

For the $\cPT$-symmetric model based on 
$\Gamma_{PT_2}=F_{\mu\nu}\gamma^\mu \gamma^\nu$, where $F_{\mu\nu}=-F_{\nu\mu}$ 
is real, the first goal is  to rationalize the full fermion propagator
$$
S(p) = (\slashed p - m-  g F_{\mu\nu}\gamma^\mu \gamma^\nu)^{-1}
$$
in order to evaluate the spinor trace in (\ref{s3e2}).
We expand $S(p)$ with 
$(\slashed p - m+  g F_{\alpha\beta}\gamma^\alpha \gamma^\beta)$. 
The denominator then takes the form
\begin{align*}
&(\slashed p - m-  g F_{\mu\nu}\gamma^\mu \gamma^\nu)\,
(\slashed p - m+  g F_{\alpha\beta}\gamma^\alpha \gamma^\beta) =\\
&
(p^2+m^2-2g^2f_1) -2m\slashed p -4g F_{\mu\nu}p^\nu\gamma^\mu 
+8 i g^2f_2\gamma_5 ,
\end{align*}
where we have used the relations
$F_{\mu\nu}(\gamma^\mu \gamma^\nu \slashed p -
\slashed p \gamma^\mu \gamma^\nu) = 4F_{\mu\nu} p^\nu \gamma^\mu$
and 
$F_{\mu\nu}F_{\alpha\beta}\gamma^\mu\gamma^\nu\gamma^\alpha\gamma^\beta
= 2f_1 \mathbbm{1} -8 i f_2 \gamma_5$, with 
\begin{align}
\label{s3e7}
f_1 &= 2(F_{01}^2+F_{02}^2+F_{03}^2-F_{12}^2-F_{13}^2-F_{23}^2) = 
\mathrm{tr}[F^2],\\
\label{s3e8}
f_2 &= (F_{01}F_{23}-F_{02}F_{13}+F_{03}F_{12}) = \mathrm{Pf}(F).
\end{align}
The denominator still contains non-scalar contributions, thus we expand $S(p)$
again with a factor that has the opposite sign in those terms:
the denominator of the resulting expression becomes
\begin{align*}
&\bigl[(p^2+m^2-2g^2f_1) -2m\slashed p -4g F_{\mu\nu}p^\nu\gamma^\mu 
+8i g^2f_2\gamma_5\bigr]\\
\times& 
\bigl[(p^2+m^2-2g^2f_1) +2m\slashed p +4g F_{\alpha\beta}p^\beta\gamma^\alpha 
-8ig^2f_2\gamma_5 \bigr] \\
&=
(p^2+m^2-2g^2 f_1)^2 -4m^2p^2 +64g^4 f_2^2 \\
& \qquad\qquad +16g^2p^\mu F_{\mu\nu}F^\nu_{\,\,\,\alpha} p^\alpha,
\end{align*}
where we utilized $\{\gamma^\mu,\gamma^\nu\}=2\eta^{\mu\nu}$, 
$\{\gamma^\mu,\gamma_5\}=0$, and $F_{\mu\nu}=-F_{\nu\mu}$.
It is now a scalar and the full propagator can then be written in the 
rationalized form
\begin{widetext}
\begin{align}
\label{s3e9}
&S(p) =
\frac{ 
(\slashed p - m+  g F_{\mu\nu}\gamma^\mu \gamma^\nu)
\bigl[(p^2+m^2-2g^2f_1) +2m\slashed p +4g F_{\alpha\beta}p^\beta\gamma^\alpha 
-8ig^2f_2\gamma_5 \bigr]
}{ 
(p^2+m^2-2g^2 f_1)^2 -4m^2p^2 +64g^4 f_2^2
+16g^2p^\mu F_{\mu\nu}F^\nu_{\,\,\,\alpha} p^\alpha
}.
\end{align}
In multiplying out the numerator, we notice that almost all resulting terms
have a vanishing spinor trace.
Using that $\mathrm{tr}[\mathbbm{1}]=4$,
$\mathrm{tr}[\gamma^\mu\gamma^\nu]=4\eta^{\mu\nu}$, and $F_\mu^{\,\,\mu}=0$ 
we find the trace of the full propagator to be
\begin{align}
\label{s3e10}
&\mathrm{tr}[S(p)] =
\frac{ 
4m (p^2-m^2+2g^2f_1)
}{ 
(p^2+m^2-2g^2 f_1)^2 -4m^2p^2 +64g^4 f_2^2
+16g^2p^\mu F_{\mu\nu}F^\nu_{\,\,\,\alpha} p^\alpha
}.
\end{align}
\end{widetext}

For the momentum integration in (\ref{s3e2}) we 
now regularize the integral in the Euclidean four-momentum cutoff method.
That is, we first change to a Euclidean system by denoting $p_0=ip_4$ and
$F_{0k}=iF_{4k}, \forall k \in [1,3]$, such that $d^4p= id^4p_E$, $p^2 = -p_E^2$ and 
$p^\mu F_{\mu\nu}F^\nu_{\,\,\,\alpha} p^\alpha 
= -p_E\cdot F_E\cdot F_E\cdot p_E$, where 
the (complex) matrix $F_E$ is
\begin{equation}
\label{s3e11}
F_E = 
\begin{bmatrix}
0 & F_{41} & F_{42} & F_{43} \\
-F_{41} & 0 & F_{12} & F_{13} \\
-F_{42} & -F_{12} & 0 & F_{23} \\
-F_{43} & -F_{13} & -F_{23} & 0
\end{bmatrix} .
\end{equation}
In this form the dependence on $F_E$ is somewhat unwieldy.
When $F_E$ is diagonalizable, it is orthogonally so
\cite{hj}, such that a transformation $Q$ exists, with $Q^TQ=\mathbbm{1}$
and $Q^T F_E\, Q = \mathrm{diag}(\lambda_1,-\lambda_1, \lambda_2,-\lambda_2)$, 
where 
\begin{align}
\label{s3e12}
\lambda_{1,2} =& \tfrac{1}{2} \sqrt{f_1\mp\sqrt{f_1^2+16f_2^2}}.
\end{align}
Thus transforming the Euclidean four-momentum according to $p_E\rightarrow~Q\cdot~p_E$ 
leaves the integral measure invariant and results in a much more convenient 
dependence of (\ref{s3e10}) on the momentum components.
This of course relies on the diagonalizability of $F_E$, which from the eigenvalues (\ref{s3e12}) we see to be so, provided that $f_2$ does not vanish.
Since $f_2$ as given in (\ref{s3e8}) implies that $f_2^2=\mathrm{Det}(F)$,
this is a minor restriction and will be considered to be the case
in what follows.
With the introduction of the four-momentum cutoff $\Lambda$, the 
regularized integral (\ref{s3e2}) over the trace (\ref{s3e10}) then has the form
\begin{widetext}
\begin{equation}
\label{s3e13}
I_{PT_2}= 
-4i m \int^\Lambda\!\! d^4 p_E \, 
\frac{ 
(p_E^2+m^2-2g^2f_1)
}{ 
(p_E^2-m^2+2g^2 f_1)^2 +4m^2p_E^2 +64g^4 f_2^2
-16g^2\lambda_1^2(p_4^2+p_1^2)
-16g^2\lambda_2^2(p_2^2+p_3^2)
}.
\end{equation}
To evaluate this, we first express the momentum components in two sets of polar coordinates, where $p_4=R_1 \cos\phi_1$, $p_1=R_1 \sin\phi_1$ and $p_2=R_2 \cos\phi_2$, $p_3=R_2 \sin\phi_2$. Equation (\ref{s3e13}) becomes
\begin{equation}
\label{s3e14}
I_{PT_2}= 
-16 i \pi^2  m \int^\Lambda\!\! dR_1 dR_2   
\frac{ 
R_1 R_2 (R_1^2+R_2^2+m^2-2g^2f_1)
}{ 
(R_1^2+R_2^2-m^2+2g^2 f_1)^2 +4m^2(R_1^2+R_2^2) +64g^4 f_2^2
-16g^2(\lambda_1^2 R_1^2+\lambda_2^2 R_2^2)
}.
\end{equation}
\end{widetext}
Now, combining $R_1$ and $R_2$ in polar coordinates $R_1 = r\cos\theta$
and $R_2 =r \sin\theta$ with $\theta \in [0,\pi/2]$ and $r\in[0,\Lambda]$
yields
\begin{align}
\label{s3e15}
\begin{split}
I_{PT_2}= &
-16 i \pi^2  m \int_0^\Lambda\!\! dr 
r^3  (r^2+m^2-2g^2f_1)\\
&\times
\int_0^{\pi/2}\!\! d\theta   
\frac{ 
\cos\theta \sin\theta
}{ 
A(r)-B(r)\cos^2\theta
},
\end{split}
\end{align}
where 
\begin{align}
\label{s3e16}
\begin{split}
A(r)=& (r^2+m^2-2g^2f_1^2)+64g^4f_2^2\\
&+4g^2r^2(f_1-\sqrt{f_1^2+16f_2^2}) ,
\end{split} \\
B(r)=&-8g^2r^2\sqrt{f_1^2+16f_2^2}.
\end{align}
The angular part is a standard integral \cite{gr},
and $I_{PT_2}$ becomes
\begin{align}
\label{s3e15}
\begin{split}
I_{PT_2}= &
 -i \pi^2  m \int_0^\Lambda\!\! dr 
\frac{r (r^2+m^2-2g^2f_1)}{g^2\sqrt{f_1^2+16f_2^2}}
\ln\Big[1-\frac{B(r)}{A(r)}\Big]
\end{split}
\end{align}
 and the remaining radial integration can then be rewritten as
\begin{align}
\label{s3e16}
\begin{split}
&I_{PT_2}= 
\frac{i \pi^2  m}{2g^2\sqrt{f_1^2+16f_2^2}}
 \int_0^{\Lambda^2}\!\! dz \,
 (z+m^2-2g^2f_1)\\
 &\times
 \bigl\{\ln(z+a_1)+\ln(z+a_2)-\ln(z+a_3)-\ln(z+a_4)\bigr\} ,
\end{split}
\end{align}
with 
\begin{align}
\label{s3e17}
a_{1,2} =& m^2-2g^2\sqrt{f_1^2+16f_2^2}\pm
\sqrt{4m^2g^2(f_1-\sqrt{f_1^2+16f_2^2})} ,\\
\label{s3e18}
a_{3,4} =& m^2+2g^2\sqrt{f_1^2+16f_2^2}\pm
\sqrt{4m^2g^2(f_1+\sqrt{f_1^2+16f_2^2})} .
\end{align}
After integration and some further simplification this then becomes
\begin{align}
\label{s3e19}
\begin{split}
I_{PT_2}=
\frac{i \Lambda^3 \pi^2  \tilde{m}}{2\tilde{g}^2\sqrt{1+f^2}}
\Bigl[&
(\tilde{a}_1+1) \ln(\tfrac{1+\tilde{a}_1}{\tilde{a}_1})
(\tilde{m}^2-2\tilde{g}^2+\tfrac{1-\tilde{a}_1}{2})\\
+&
(\tilde{a}_2+1) \ln(\tfrac{1+\tilde{a}_2}{\tilde{a}_2})
(\tilde{m}^2-2\tilde{g}^2+\tfrac{1-\tilde{a}_2}{2})\\
-&
(\tilde{a}_3+1) \ln(\tfrac{1+\tilde{a}_3}{\tilde{a}_3})
(\tilde{m}^2-2\tilde{g}^2+\tfrac{1-\tilde{a}_3}{2})\\
-&
(\tilde{a}_4+1) \ln(\tfrac{1+\tilde{a}_4}{\tilde{a}_4})
(\tilde{m}^2-2\tilde{g}^2+\tfrac{1-\tilde{a}_4}{2})\\
-&
4\tilde{g}^2\sqrt{1+f^2}
\Bigr]
\end{split}
\end{align}
in terms of the rescaled quantities $\tilde{m}=m \Lambda^{-1}$, 
$\tilde{a}=a\Lambda^{-2}$, $\tilde{g}^2 = g^2 f_1 \Lambda^{-2}$, and 
$f=4f_2/f_1$.
Note that $f$ is a real valued parameter.
When we consider the limit of vanishing bare mass $m_0$, denote 
$\tilde{G}=G\Lambda^2$, and use the result for the momentum integral 
$I_{PT_2}$, the general gap equation (\ref{s3e1}) takes the algebraic form
\begin{figure*}
\centering
\subfloat[]{
\includegraphics[width=0.32\textwidth]
{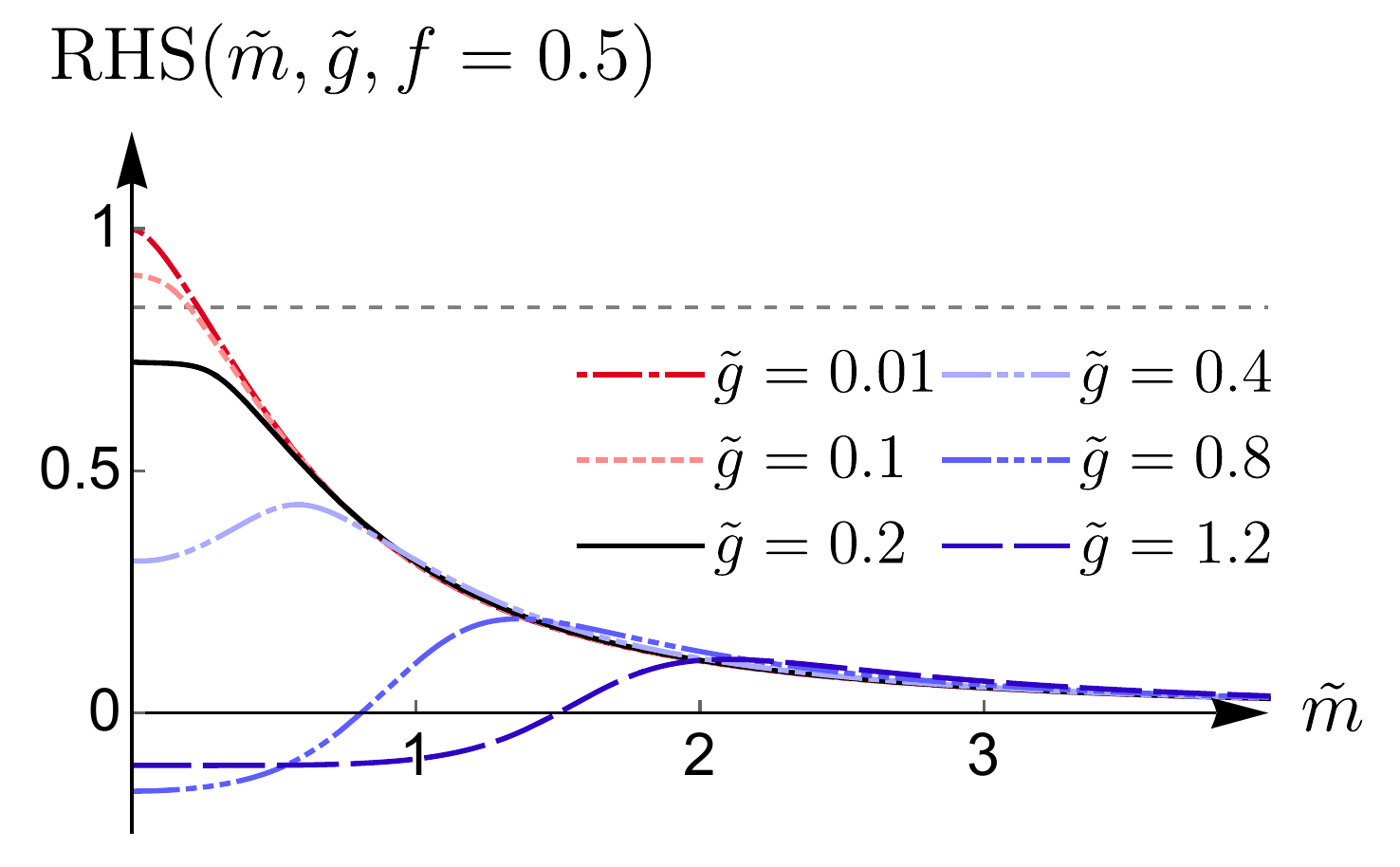}
\label{f2}
}
\hfill
\subfloat[]{
\includegraphics[width=0.32\textwidth]
{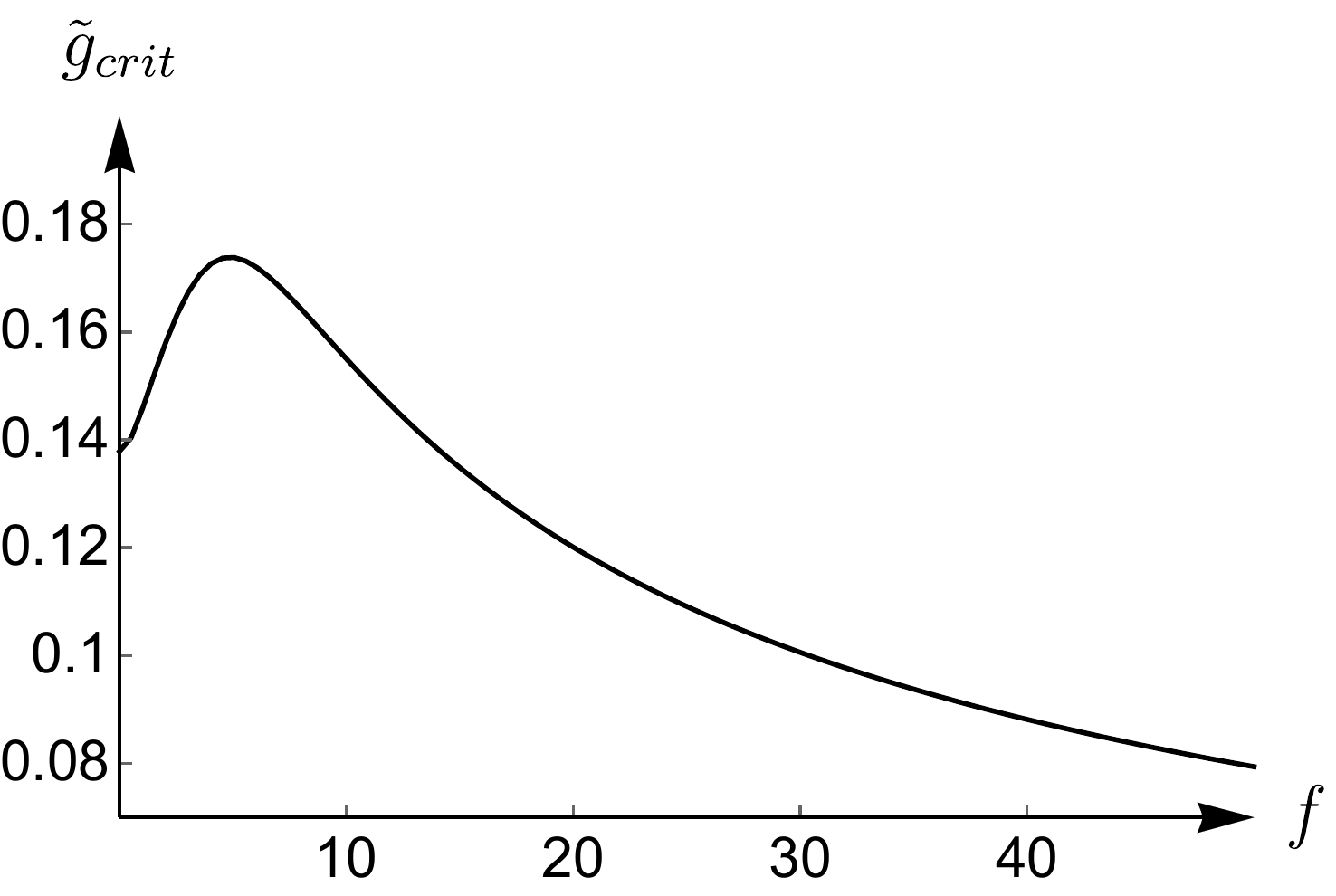}
\label{f3}
}
\hfill
\subfloat[]{
\includegraphics[width=0.32\textwidth]
{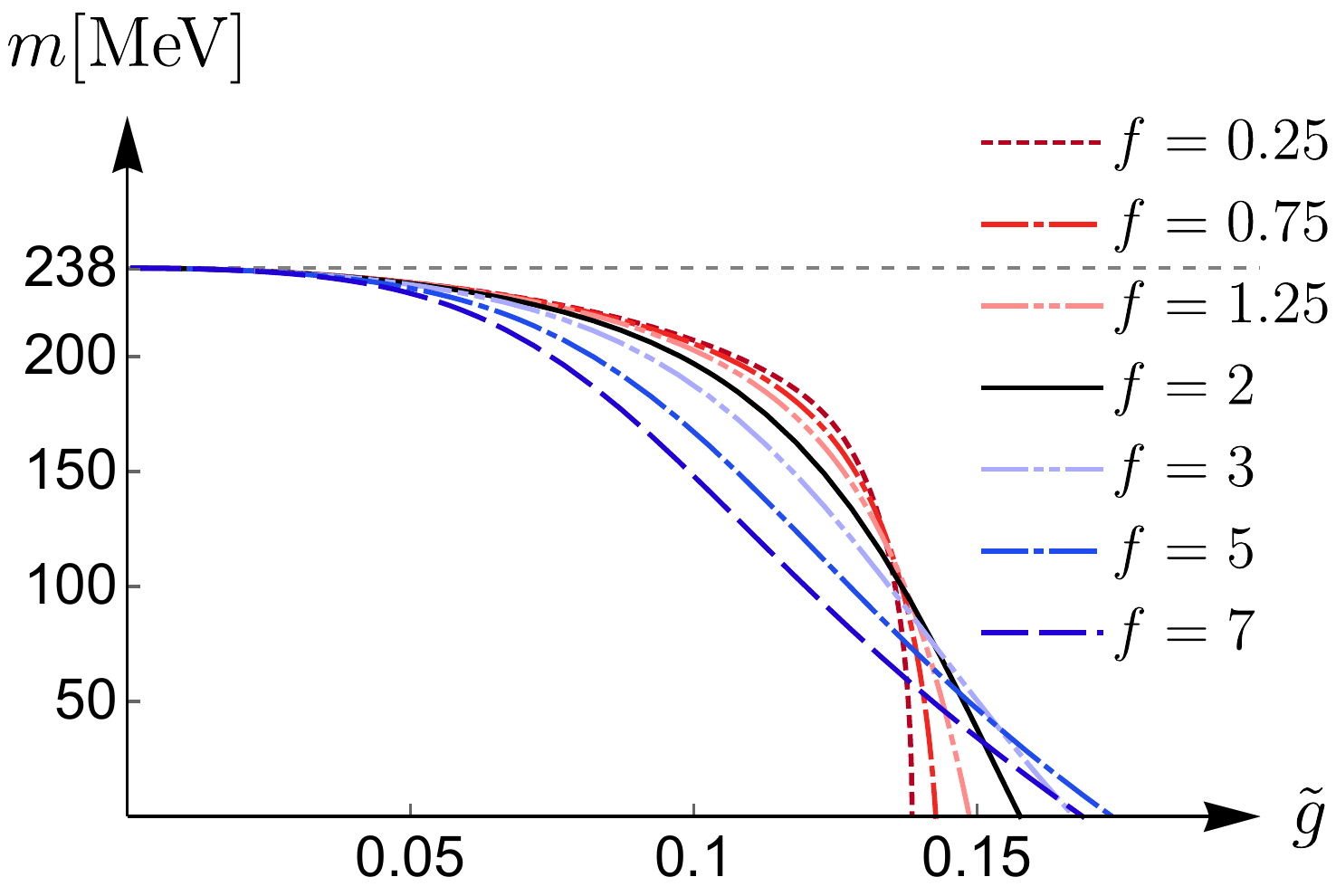}
\label{f4}
}
\caption{
(a)
Behavior of the right hand side of the gap equation (\ref{s3e20}) based on 
$\Gamma_{PT_2}$ as a function of the mass $\tilde{m}$ and for different values of 
the coupling $\tilde{g}$. 
The value for $f=0.5$ is fixed, but this case is representative of the general 
behavior. 
The horizontal line represents the left hand side of the gap equation.
(b)
Behavior of the critical coupling value $\tilde{g}_{crit}$ up to which the 
gap equation (\ref{s3e20}) has real mass solutions as a function of 
$f$.
(c)
Behavior of the effective fermion mass $m$ in MeV as a
function of the coupling constant up to the corresponding critical values for 
different values of $f$.
}
\end{figure*}
\begin{widetext}
\begin{align}
\label{s3e20}
\frac{2 \pi^2}{\tilde{G} N_c N_f} = \frac{1}{8 \tilde{g}^2\sqrt{1+f^2}}
\Big\{&
4\tilde{g}^2\sqrt{1+f^2}
-
(\tilde{a}_1+1) \ln(\tfrac{1+\tilde{a}_1}{\tilde{a}_1})
(\tilde{m}^2-2\tilde{g}^2+\tfrac{1-\tilde{a}_1}{2})
-
(\tilde{a}_2+1) \ln(\tfrac{1+\tilde{a}_2}{\tilde{a}_2})
(\tilde{m}^2-2\tilde{g}^2+\tfrac{1-\tilde{a}_2}{2})\notag \\
&+
(\tilde{a}_3+1) \ln(\tfrac{1+\tilde{a}_3}{\tilde{a}_3})
(\tilde{m}^2-2\tilde{g}^2+\tfrac{1-\tilde{a}_3}{2})
+
(\tilde{a}_4+1) \ln(\tfrac{1+\tilde{a}_4}{\tilde{a}_4})
(\tilde{m}^2-2\tilde{g}^2+\tfrac{1-\tilde{a}_4}{2})
\Big\}
\end{align}
\end{widetext}
in the limit of vanishing bare mass $m_0$. 
We note that in the limit $\tilde{g} \to 0$ one recovers the known gap 
equation of the conventional Hermitian NJL model in this regularization 
scheme \cite{spk}:
\begin{equation}
\label{s3e21}
\frac{2 \pi^2}{\tilde{G} N_c N_f} = 1 - \tilde{m}^2 
\ln\Big(\frac{1+\tilde{m}^2}{\tilde{m}^2} \Big) .
\end{equation}
The behavior of the right hand side of (\ref{s3e20}) is shown in Fig.~\ref{f2}
as a function of $\tilde{m}$ and for different values of $\tilde{g}$. The 
parameter $f=0.5$ is fixed, but is representative of the general behavior.
The horizontal line denotes the constant real positive left hand side 
of (\ref{s3e20}) for $\tilde{G}=3.93$, $N_c=3$ and $N_f=2$. 
For sufficiently small coupling values of $\tilde{g}$ an intersection, and 
therefore a real fermion mass solution $\tilde{m}$ to the gap equation, can always be 
found. 
However, beyond a critical value $\tilde{g}_{crit}$ the maximum of the 
right hand side no longer exceeds the constant given by the left hand side 
and real solutions no longer exist. 
The value of this critical coupling depends on the parameter $f$ and its 
behavior is shown in Fig.~\ref{f3}. 
It approaches zero asymptotically as $f$ increases to infinity.
Figure~\ref{f4} displays the behavior of the fermion mass solution $m$ (in
MeV) as a function of $\tilde{g}$ up to the corresponding
critical couplings for different values of $f$.
We note that, independent of $f$, the effective mass {\it decreases} with increasing
coupling $\tilde{g}$. Thus a current quark mass can not 
be generated dynamically in this model. 

Altogether, we find that - similar to the model discussed in 
Sec.~\ref{s3a} - this $\cPT$-symmetric modified (non-Hermitian) NJL model admits a 
finite region of unbroken $\cPT$ symmetry, even for vanishing bare mass $m_0$. 
However, while real fermion mass solutions do exist for coupling values
up to the critical value $\tilde{g}_{crit}$, these solutions always describe an 
effective mass loss; an {\it additional} dynamical mass generation over and above that obtained through the 
four-fermion interaction is not possible.

\subsection{$\Gamma_{aPT_1}=iB_\mu \gamma^\mu$}
\label{s3c}

The case of the non-$\cPT$-symmetric model based on including the vector background field $\Gamma_{aPT_1}=iB_\mu \gamma^\mu$ is structurally very similar to the 
$\cPT$-symmetric one  discussed in Sec.~\ref{s3a}. It is also the 
only other non-Hermitian, bilinear extension of the NJL model that preserves
chiral symmetry, as discussed in Sec.~\ref{s2}.

To find the effective fermion mass of the model, we first
rationalize the full propagator
$$
S(p) = (\slashed p - m-  igB_\mu \gamma^\mu)^{-1} ,
$$
finding that it can be written as 
\begin{equation}
\label{s3e23}
S(p) =
\frac{ 
\slashed p + m-  iB_\mu \gamma^\mu
}{ 
p^2-m^2-g^2B^2 -2ig B_\mu p ^\mu
},
\end{equation}
from which the spinor trace is readily obtained:
\begin{figure*}
\centering
\null
\subfloat[]{
\includegraphics[width=0.48\textwidth]
{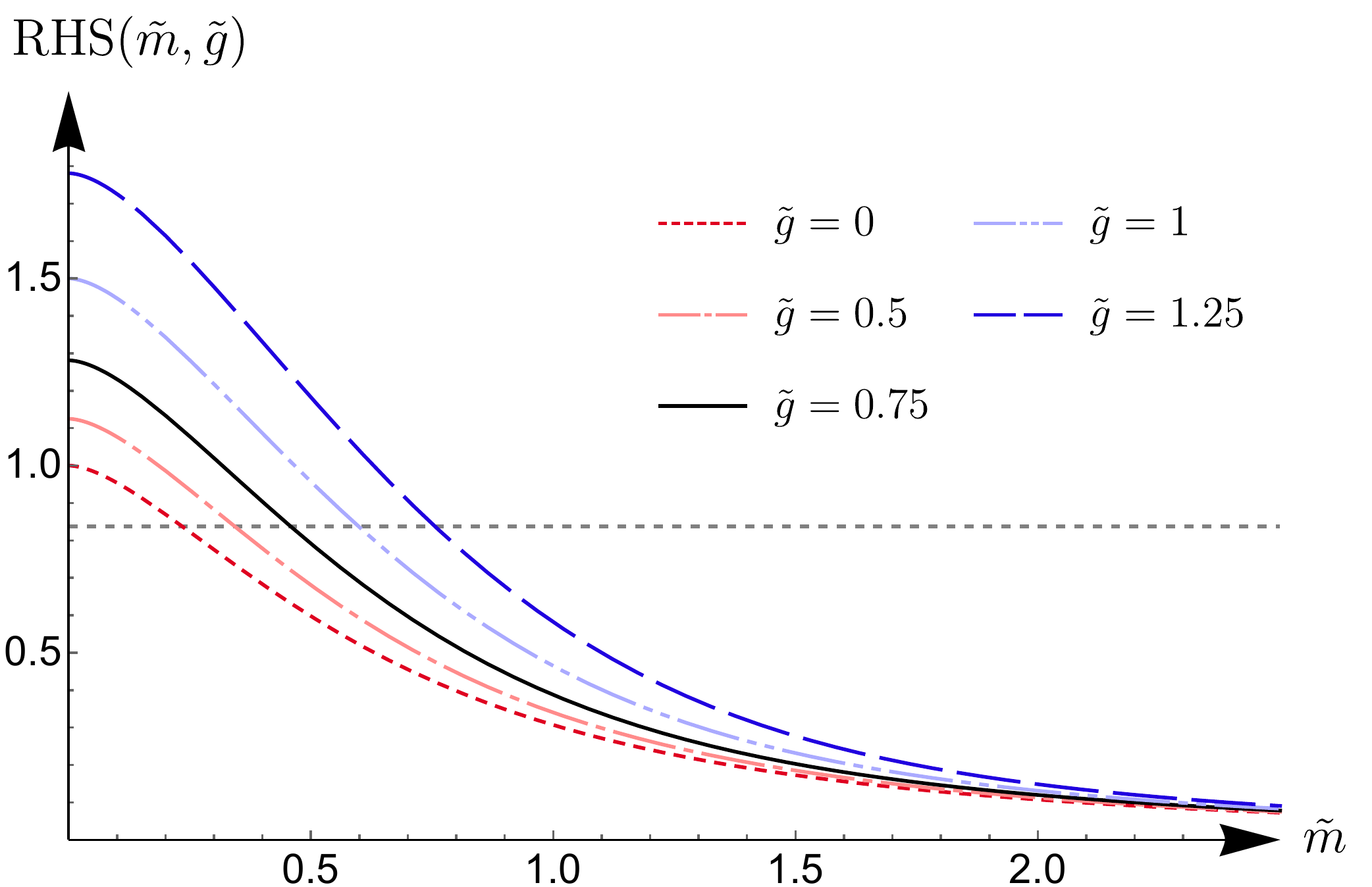}
\label{f5}
}
\hfill
\subfloat[]{
\includegraphics[width=0.48\textwidth]
{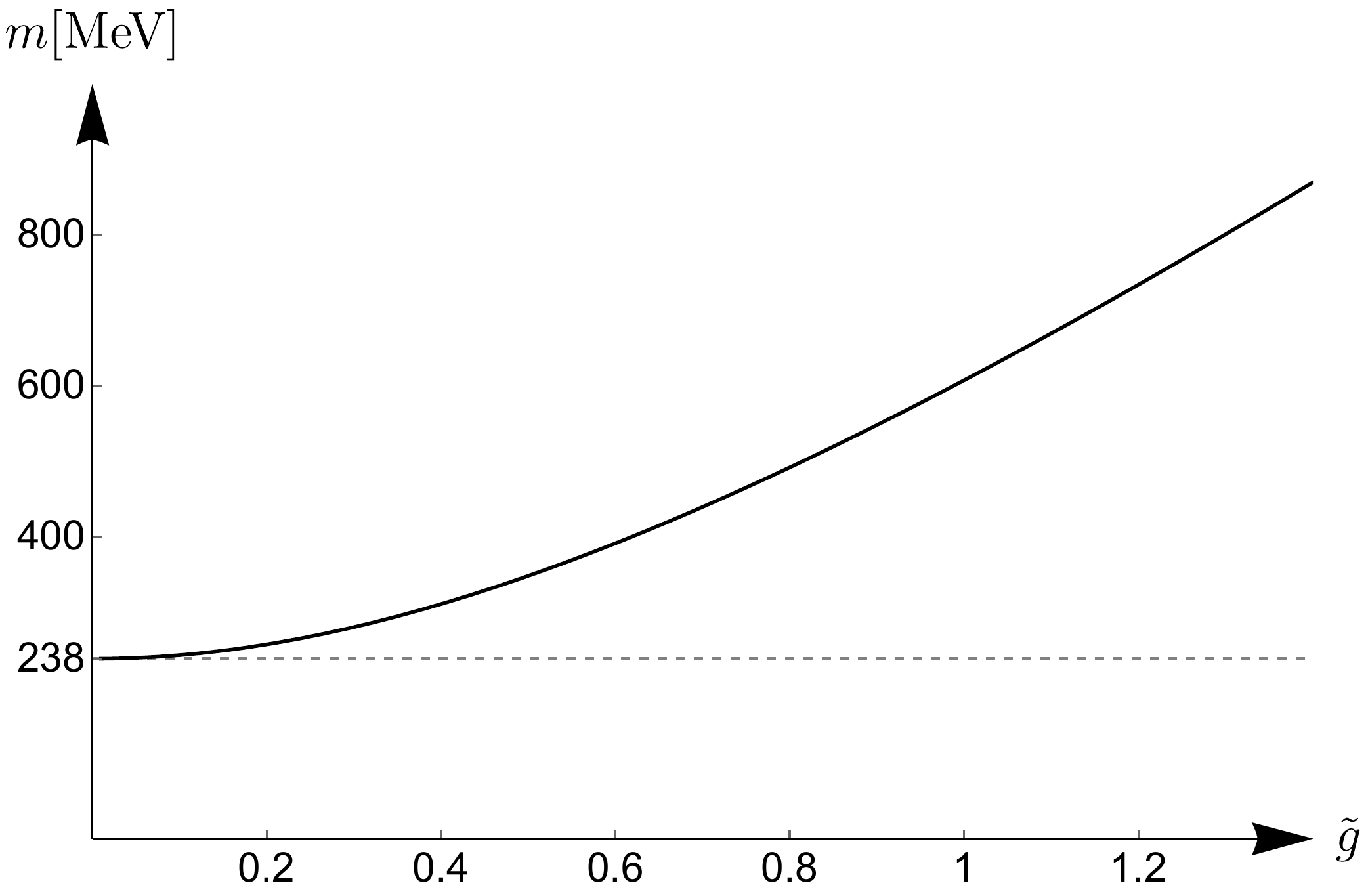}
\label{f6}
}
\caption{ 
In (a) the behavior of the right hand side of the gap equation (\ref{s3e28}) 
based on $\Gamma_{aPT_1}$ is shown as a function of the mass $\tilde{m}$ for different values of the coupling $\tilde{g}$. 
The horizontal line represents the left hand side of the gap equation.
Figure (b) displays the behavior of the real mass solutions $m$ in MeV as
function of the coupling constant.}
\end{figure*}
\begin{align}
\label{s3e24}
&\mathrm{tr}[S(p)] =
\frac{ 
4m 
}{ 
p^2-m^2-g^2B^2 -2ig B_\mu p ^\mu
}.
\end{align}
Integrating out the four-momentum dependence of this expression results 
in an algebraic form of the gap equation of the model, cf. Eq.~(\ref{s3e2}).
As described previously in Sec.~\ref{s3a}, we evaluate this integral in the 
Euclidean four-momentum cutoff regularization scheme:
The cutoff $\Lambda$ is introduced after transforming to
coordinates $p_0=ip_4$ and $B_0=iB_4$, such that $p^2=-p_E^2$, 
$B^2=-B_E^2$, and 
$B_\mu p^\mu = -B_E\cdot p_E$.
Thus
$$
I_{aPT_1}= 
\int^\Lambda\!\! d^4 p_E \, 
\frac{ 
-4i m 
}{ 
p_E^2+m^2-g^2B_E^2 -2ig(B_E \cdot p_E)
} .
$$
In spherical coordinates with the zenith direction along $B_E$, the Euclidean 
scalar product $B_E \cdot p_E = \lvert B_E \rvert r \cos\theta$
contains only the radius $r=\lvert p_E\rvert$ and the zenith angle $\theta$, 
so that we can write
\begin{equation}
\label{s3e27}
I_{aPT_1}= 
\frac{8 \pi m}{g\lvert B_E\rvert}
\int_0^\Lambda\!\! dr  \, r^2
\int_0^\pi \! d\theta  \,
\frac{\sin^2 \theta}{A(r)+\cos\theta} ,
\end{equation}
where $A(r)=(r^2+m^2-g^2\lvert B_E\rvert^2)/(-2igr\lvert B_E\rvert)$.
Both the angular and the resulting radial integrals are standard \cite{gr}.
In terms of the rescaled quantities $\tilde{m}=m\Lambda^{-1}$ and 
$\tilde{g}=g \lvert B_E\rvert \Lambda^{-1}$,
the latter of which is proportional to the amplitude of the vector background field, (\ref{s3e27}) becomes:
\begin{widetext}
\begin{align}
\label{s3e28}
\begin{split}
I_{aPT_1}=
\frac{i \Lambda^3 \pi^2  \tilde{m}}{\tilde{g}^2}
\Bigl\{&
1+(\tilde{m}^2-\tilde{g}^2)(2+\tilde{m}^2+\tilde{g}^2)
-
(1+\tilde{m}^2+\tilde{g}^2 )
\sqrt{(1 + \tilde{m}^2 + \tilde{g}^2)^2 - 4 \tilde{g}^2 \tilde{m}^2}
\\
&+
4\tilde{m}^2\tilde{g}^2 \ln\Big[ \frac{1}{2 \tilde{m}^2} 
\Bigl( \sqrt{(1 + \tilde{m}^2 + \tilde{g}^2)^2 - 4 \tilde{g}^2 \tilde{m}^2} 
+ 1 + \tilde{m}^2 + \tilde{g}^2 \Bigr) \Big] 
\Bigr\} .
\end{split}
\end{align}
Denoting $\tilde{G}=G\Lambda^2$, the general gap equation (\ref{s3e2}) in the limit of vanishing bare mass $m_0$ thus
takes the algebraic form
\begin{align}
\label{s3e28}
\begin{split}
\frac{2 \pi^2}{\tilde{G} N_c N_f} =
\frac{1}{4\tilde{g}^2}
\Bigl\{&
(1+\tilde{m}^2+\tilde{g}^2 )
\sqrt{(1 + \tilde{m}^2 + \tilde{g}^2)^2 - 4 \tilde{g}^2 \tilde{m}^2}
-1-(\tilde{m}^2-\tilde{g}^2)(2+\tilde{m}^2+\tilde{g}^2)
\\
&-
4\tilde{m}^2\tilde{g}^2 \ln\Big[ \frac{1}{2 \tilde{m}^2} 
\Bigl( \sqrt{(1 + \tilde{m}^2 + \tilde{g}^2)^2 - 4 \tilde{g}^2 \tilde{m}^2} 
+ 1 + \tilde{m}^2 + \tilde{g}^2 \Bigr) \Big] 
\Bigr\}.
\end{split}
\end{align}
\end{widetext}
The gap equation (\ref{s3e21}) of the standard NJL model
is recovered in the limit $\tilde{g} \to 0$.

The right hand side of (\ref{s3e28}) is a real function 
of the scaled mass $\tilde{m}$ and coupling constant $\tilde{g}$. 
Its behavior is shown in Fig.~\ref{f5}. 
Intersections with the real constant left hand side, visualized by a horizontal 
line, can always be found when evaluated within the context of QCD.
The corresponding fermion mass solution $m$ (in MeV) is shown 
in Fig.~\ref{f6} as a function of the coupling constant $\tilde{g}$.  
Note that, contrary to the $\cPT$-symmetric models discussed in  \ref{s3a} and 
\ref{s3b}, real mass solutions are not limited to a finite coupling constant 
regime and increase monotonously with the value of the coupling constant.
A current quark mass is generated dynamically in this model: 
the equivalent of a bare up quark mass, $m_u=(1.7-3.3)$~MeV,
requires coupling values in the range 
$\tilde{g}\approx(0.059-0.083)$
and for the bare down quark mass,
$m_d=(4.1-5.8)$~MeV, coupling values  
$\tilde{g}\approx(0.092-0.110)$
are necessary.

Altogether, we find that real fermion mass solutions are obtained for
all values of the coupling constant $\tilde{g}$ in this chirally symmetric non-Hermitian 
extension of the NJL model, even though $\cPT$ symmetry is broken explicitly. 
Moreover, the generation of an additional dynamical mass mimicking a bare fermion mass occurs.

\subsection{$\Gamma_{aPT_2}=\gamma_5$}
\label{s3d}

\begin{figure}
\centering
\includegraphics[width=0.48\textwidth]
{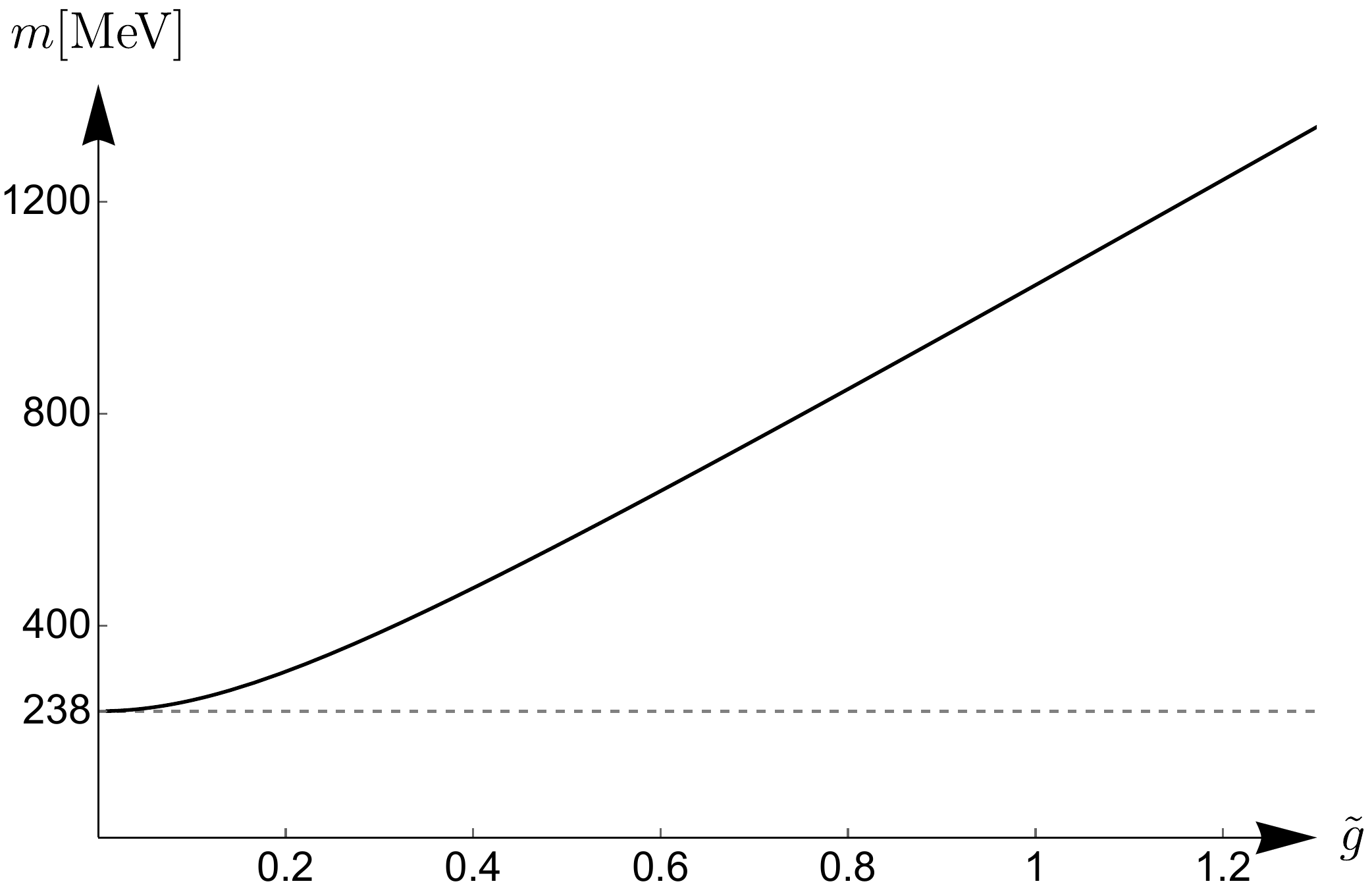}
\caption{ 
\label{f7}
Behavior of the fermion mass solution $m$ of the modified NJL model based on $\Gamma_{aPT_2}$ 
as a function of the scaled coupling constant 
$\tilde{g}=g\Lambda^{-1}$.
}
\end{figure}

The analysis of the non-$\cPT$-symmetric model based on 
the pseudoscalar term $\Gamma_{aPT_2}= \gamma_5$ is straight forward:
The full fermion propagator  
$$
S(p) = (\slashed p - m-  g\gamma_5)^{-1}
$$
can be rationalized and simplified to the form
\begin{equation}
\label{s3e30}
S(p) =
\frac{ 
\slashed p + m-  g\gamma_5
}{ 
p^2-m^2+g^2
},
\end{equation}
which has the spinor trace
\begin{align}
\label{s3e31}
&\mathrm{tr}[S(p)] =
\frac{ 
4m 
}{ 
(p^2-m^2+g^2)
}.
\end{align}
Since this function only depends on the square of the four-momentum, 
the momentum integral (\ref{s3e2}) can be evaluated directly: in spherical Euclidean coordinates, 
with $p^2=-p_E^2$, we introduce the Euclidean four-momentum cutoff
$\Lambda$ as the bound of the radial integration and subsequently find that  
\begin{equation}
\label{s3e32}
I_{aPT_2}=
4i \Lambda^3 \pi^2 \tilde{m} \Big[
(\tilde{m}^2-\tilde{g}^2) \ln\big(\frac{1+\tilde{m}^2-\tilde{g}^2}
{\tilde{m}^2-\tilde{g}^2} \big)
-1 \Big],
\end{equation}
in terms of the rescaled parameters $\tilde{m}=m\Lambda^{-1}$ and 
$\tilde{g}=g\Lambda^{-1}$.
The gap equation (\ref{s3e1}) of this model thus takes the algebraic form 
\begin{equation}
\label{s3e33}
\frac{2 \pi^2}{\tilde{G} N_c N_f} = 1 - (\tilde{m}^2-\tilde{g}^2) 
\ln\big(\frac{1+\tilde{m}^2-\tilde{g}^2}{\tilde{m}^2-\tilde{g}^2} \big) 
\end{equation}
in the limit $m_0\rightarrow 0$, which reduces to the gap equation (\ref{s3e21}) of the standard NJL model in the limit of vanishing 
coupling constant values $\tilde{g}$.

Equation (\ref{s3e33}) can be solved algebraically in terms of the  
Lambert $W$ function \cite{cghjk}. We find the real fermion mass solution
\begin{equation}
\label{s3e34}
\tilde{m}= \sqrt{\tilde{g}^2+[\tfrac{1}{c}W_{-1}(c e^c)-1]^{-1}}
=  \sqrt{\tilde{g}^2+\tilde{m}_{NJL}^2},
\end{equation}
where, in the context of QCD, $c = 2\pi^2/\tilde{G}N_cN_f-1$ is a negative real constant and 
$\tilde{m}_{NJL}$ is the (scaled) fermion mass of the standard NJL model.
The behavior of this solution is shown in Fig.~\ref{f7}.
Similar to the case discussed in the previous Sec.~\ref{s3c}, we find that 
the existence of real fermion mass solutions is, in 
contrast to the $\cPT$-symmetric models, not limited to a finite coupling constant regime and the mass increases again monotonously with the coupling
constant value $\tilde{g}$.
A current quark mass equivalent to that of a bare up quark mass, 
$m_u=(1.7-3.3)$~MeV, is dynamically generated for coupling values in the range 
$\tilde{g}\approx(0.028-0.039)$
and for the equivalent of a the bare down quark mass,
$m_d=(4.1-5.8)$~MeV, coupling values  
$\tilde{g}\approx(0.044-0.052)$ are required.

We remark that the analytical mass solution (\ref{s3e34}) coincides with 
that found for the renormalized, modified Gross-Neveu (GN) model in 
\cite{fbk}, which can be seen as a $1+1$ dimensional version of the NJL model. 
We emphasize that while the pseudoscalar term $\Gamma_{aPT_2}=\gamma_5$ does not give
rise to a $\cPT$-symmetric bilinear in $3+1$ dimensions, its $1+1$ dimensional 
equivalent is, in fact, the only possible $\cPT$-symmetric, non-Hermitian,
bilinear extension of the GN model \cite{bkb,fbk}.

Altogether, we find that the fermion mass solution of this model
resembles that found in the previous Sec.~\ref{s3c}: 
even though $\cPT$ symmetry is explicitly broken, real mass solutions are found for all coupling constant values and an {\it additional} dynamical mass generation, over and above
that generated by the standard NJL model, occurs. 
In contrast to the model based on $\Gamma_{aPT_1}$, this model does, however, 
not preserve chiral symmetry.

\subsection{$\Gamma_{aPT_3}=i\mathbbm{1}$}
\label{s3e}

Modifying the NJL model by adding a bilinear term based on the scalar 
$\Gamma_{aPT_3}= i\mathbbm{1}$
amounts to including an imaginary mass contribution 
proportional to the coupling constant $g$ into the propagator:
$$
S(p) = (\slashed p - m-  ig)^{-1}.
$$
Correspondingly, the gap equation (\ref{s3e1}) is obtained as that of the 
standard NJL model with an additional contribution to the mass terms arising in the momentum integral on the right hand side. In the four-momentum cutoff regularization scheme it reads
\begin{equation}
\label{s3e36}
\frac{2 \pi^2}{\tilde{G} N_c N_f} =
\frac{\tilde{m}+i\tilde{g}}{\tilde{m}}
\Bigl\{1 - (\tilde{m}+i\tilde{g})^2 
\ln\Big[\frac{1+(\tilde{m}+i\tilde{g})^2}{(\tilde{m}+i\tilde{g})^2}\Big] 
\Bigr\}
\end{equation}
in terms of the rescaled parameters $\tilde{m}=m\Lambda^{-1}$ and 
$\tilde{g}=g\Lambda^{-1}$.
For a vanishing coupling constant $\tilde{g}$ this simplifies to the gap 
equation (\ref{s3e21}) of the standard NJL model in this 
regularization scheme.
Otherwise, the right hand side of (\ref{s3e36}) takes on inherently complex
values and an intersection with the purely real left hand side, and thus a 
real fermion mass solution can not be found.

The modified NJL model based on $\Gamma_{aPT_3}$ is thus the only one of the 
non-Hermitian systems considered here that does not admit real fermionic mass solutions at all. 

\emph{In summary}:
For $\cPT$-symmetric theories  based on $\Gamma_{PT_1}$ and $\Gamma_{PT_2}$ that symmetry was realized exclusively as a broken symmetry phase in the underlying modified free Dirac models, generally giving 
rise to  complex fermion masses \cite{bkb}.
However, the addition of two-body interactions via the associated
modified NJL models restores this symmetry, leading to a finite region where $\cPT$  is unbroken. This is manifested through 
the real fermion mass solutions found in both cases (cf. \ref{s3a} and \ref{s3b}).
Nevertheless, an {\it additional} dynamical mass generation is only possible in the model 
extended by a pseudovector background field $\Gamma_{PT_1}$ (\ref{s3a}), and in this case only for
specific values of the coupling strength $\tilde g$.

In contrast, it is suprising that one can find real fermion masses in the non-Hermitian non-$\cPT$-symmetric models that include $\Gamma_{aPT_1}$ and $\Gamma_{aPT_2}$ 
at all, cf. \ref{s3c} and \ref{s3d}, 
as these can not be attributed to an unbroken symmetry phase of any kind.
Furthermore, in both models these real mass solutions are not restricted to finite coupling regions, but are obtained for all possible values of the coupling constant
$\tilde g$. 
In addition, these solutions always allow for increasing dynamically generated masses, over and above those obtained from the standard NJL model.
In the model containing $\Gamma_{aPT_3}$ in \ref{s3e} no real fermion masses were found.

Thus, the effect of a small bare mass $m_0$ in the standard NJL model 
can be mimicked by including any of the three non-Hermitian bilinear
terms $\Gamma_{PT_1}$, $\Gamma_{aPT_1}$ or $\Gamma_{aPT_2}$ into the NJL model.

\section{Meson masses}
\label{s4}

We have identified four non-Hermitian bilinear extensions of the NJL model
that give rise to purely real fermion mass solutions, three of which allow 
for an additional dynamical generation of fermion mass as a function of the coupling strength $\tilde g$.
In this section we investigate the masses of the corresponding scalar and 
pseudoscalar bound states for these three models. In the context of QCD these correspond to the $\sigma$ and $\pi$ mesons respectively.
First we recall the established self-consistent mass equations for these 
bound states and how they can be obtained from the effective meson 
interaction, cf.~\cite{njl, spk, fw}.
In Sec.~\ref{s4a} we call to mind the resulting meson mass solutions 
for the standard NJL model, followed by a discussion of the 
results found for the modified NJL model based on including the pseudoscalar term 
$\Gamma_{aPT_2}=\gamma_5$, in which both chiral and $\cPT$ symmetry are broken explicitly, c.f. Sec.~\ref{s4b}.
In Secs.~\ref{s4c} and \ref{s4d}, the scalar and pseudoscalar meson masses are analyzed
for the modified NJL models based on the vector background field $\Gamma_{aPT_1}=iB_\mu \gamma^\mu$ and the pseudovector background field
$\Gamma_{PT_1}=i\gamma_5 B_\mu \gamma^\mu$, both of which are chirally symmetric.
 
Following the discussions in \cite{spk} and \cite{fw} for the standard 
NJL model, one must contruct the effective meson interaction 
$U_{\alpha \alpha', \beta \beta'}$, where all relevant degrees of freedom are subsumed in the indices $\alpha, \alpha^\prime, \beta, \beta^\prime$. The effective interaction is expressed in terms of the (proper) polarization insertion $\Pi_{\lambda \lambda', \mu \mu'}$:
\begin{align}
\label{s4e1}
\begin{split}
iU_{\alpha \alpha', \beta \beta'}(k)&=iU^0_{\alpha \alpha', \beta \beta'}(k)\\
&+
iU^0_{\alpha \alpha', \lambda \lambda'}(k)
[-i\Pi_{\lambda \lambda', \mu \mu'}(k)]
iU_{\mu \mu', \beta \beta'}(k) .
\end{split}
\end{align} 
Here $U^{0}_{\alpha \alpha', \beta \beta'}$ denotes the bare four-point 
interaction of the model, which (in position space) has the form 
\begin{align}
\label{s4e2}
\begin{split}
U^0_{\alpha \alpha', \beta \beta'}(x,y) &=2G \delta^{(4)}(x-y) \\ 
& \times
[\delta_{\alpha \alpha'}\delta_{\beta \beta'}
+(i\gamma_5 \vec{\tau})_{\alpha \alpha'} (i\gamma_5 \vec{\tau})_{\beta \beta'}
].
\end{split}
\end{align}
Choosing
$U^{0}_{\alpha \alpha', \beta \beta'}(k)= U^{0}(k) \delta_{\alpha \alpha'} 
\delta_{\beta \beta'}$
or 
$U^{0}_{\alpha \alpha', \beta \beta'}(k)= U^{0}(k)(i\gamma_5 
\vec{\tau})_{\alpha \alpha'} (i\gamma_5 \vec{\tau})_{\beta \beta'}$
(\ref{s4e1}) simplifies to
\begin{equation}
\label{s4e3}
iU(k)=
iU^0(k)+
iU^0(k)
[-i\Pi^{s/ps}(k)]
iU(k) ,
\end{equation}
where 
$\Pi^{s}(k) = \delta_{\lambda \lambda'} \Pi^{s}_{\lambda 
\lambda', \mu \mu'}(k) \delta_{\mu \mu'}$
and
$\Pi^{ps}(k) = (i\gamma_5 \vec{\tau})_{\lambda \lambda'} \Pi^{ps}_{\lambda 
\lambda', \mu \mu'}(k) (i\gamma_5 \vec{\tau})_{\mu \mu'}$ 
for the scalar and pseudoscalar case respectively.
Equation~(\ref{s4e3}) is a geometric progression that can be 
summed to the form
$$
iU(k)= \frac{iU^0(k)}{1-2G\Pi^{s/ps}(k)}.
$$
The pole of this expression corresponds to that of the general scalar or pseudoscalar mode propagator, that is it occurs when $k^2=m_{s/ps}^2$.
Thus, in order to determine the mass $m_{s/ps}$, we have to solve the 
equation 
\begin{equation}
\label{s4e5}
1-2G\Pi^{s/ps}(k) = 0.
\end{equation}

The lowest order polarization insertion is a
closed fermion loop. One can improve this approximation by implementing 
self-consistency, 
i.e. we replace the free fermion propagator $S^0$ by the full propagator $S$.
Evaluating this contribution with the corresponding vertex functions gives
\begin{align}
\label{s4e6}
-i\Pi^{s}(k)=& 
N_c N_f \int \!\!\frac{d^4p}{(2\pi)^4} \mathrm{tr}[ S(p+k) S(p)] 
\end{align}
and similarly
\begin{equation}
\label{s4e7}
-i\Pi^{ps}(k)=
- N_c N_f \int\!\! \frac{d^4p}{(2\pi)^4} \mathrm{tr}
[\gamma_5 S(p+k) \gamma_5 S(p)]
\end{equation}
for each pseudoscalar channel.

Equation~(\ref{s4e5}) at $k^2 = m_{s/ps}^2$, 
together with (\ref{s4e6}) and (\ref{s4e7}), thus determines the mass of 
the scalar and pseudoscalar bound states. 
To simplify notation, we use the general form of the gap equation~(\ref{s3e1}), as discussed in 
Sec.~\ref{s3}, in the limit of vanishing bare mass $m_0$,  and rewrite
Eq.~(\ref{s4e5}) as:
\begin{equation}
\label{s4e8}
R^{s/ps}(p,k) - I/m = 0 \quad \text{at} \,\,\,  k^2 = m_{s/ps}^2 , 
\end{equation}
where 
\begin{align}
\label{s4e9}
R^s(k)&= \int \!\! d^4p \,\mathrm{tr}[ S(p+k) S(p)] ,\\
\label{s4e10}
R^{ps}(k)&= -\int \!\! d^4p \,\mathrm{tr}[\gamma_5 S(p+k)\gamma_5 S(p)], 
\end{align}
and $I$ defined in (\ref{s3e2}) is evaluated through
\begin{equation}
\label{s4e10-2}
\frac{I}{m}=\, \frac{1}{2m}\int \!\! d^4 p \, \bigl( {\rm tr}[S(p)] + {\rm tr}[S(p+k)]
\bigr).
\end{equation} 

As was the case for the gap equation, the equation for the self-consistent meson masses (\ref{s4e8}) of the standard NJL model
retains the same general structure for the modified NJL models
constructed with the Hamiltonian (\ref{s2e2}), since the influence of the additional bilinear terms
is accounted for in the full fermion propagator $S(p)$, cf.~\cite{fbk}.
Similar to the discussion of the gap equation in Sec.~\ref{s3}, our 
analysis of the meson masses thus relies on the evaluation of spinor traces
and momentum integrations in (\ref{s4e9}) and (\ref{s4e10}) involving the appropriate fermion propagators
of the modified NJL models.

\subsection{Standard NJL model}
\label{s4a}

In the standard NJL model the evaluation of the self-consistent meson mass equation (\ref{s4e8}) is based on the full Dirac fermion propagator
$$
S(p) = (\slashed p - m)^{-1} = \frac{\slashed p + m}{ p^2-m^2} ,
$$
with $m=m_{NJL}$ being the effective fermion mass determined by the gap equation 
(\ref{s3e1}) in the chiral limit.
From this, the spinor traces in Eqs.~(\ref{s4e9})-(\ref{s4e10-2}) are readily calculated, leading to the expressions
$$
R^{s}(k) =\int \!\! d^4p \frac{4 (p^2 + p_\mu k^\mu +m^2)
 }{ ((p+k)^2-m^2)(p^2-m^2)}
$$
and 
$$
R^{ps}(k) =\int \!\! d^4p \frac{4 (p^2 + p_\mu k^\mu -m^2)
 }{ ((p+k)^2-m^2)(p^2-m^2)},
$$
as well as 
$$
\frac{I}{m}= \int \!\! d^4 p \, \frac{4(p^2 + p_\mu k^\mu -m^2) +2 k^2 
 }{ ((p+k)^2-m^2)(p^2-m^2)}.
$$
Equation (\ref{s4e8}) then takes the form of the well-known conditions
\begin{align}
\label{s4e15}
0=&\, (k^2-4m^2) \int \!\! d^4p \frac{1}{ ((p+k)^2-m^2)(p^2-m^2)} ,\\
\label{s4e16}
0=&\, k^2 \int \!\! d^4p \frac{1}{ ((p+k)^2-m^2)(p^2-m^2)}
\end{align}
at $k^2 = m_{s/ps}^2$ for the scalar and pseudoscalar meson masses 
respectively, cf. \cite{spk}.

These conditions give rise to the apparent solutions that 
$m_{s}^2 = 4m^2 = 4m_{NJL}^2$ and $m_{ps}^2=0$, the latter of 
which is the Nambu-Goldstone mode in this model, in which 
chiral symmetry is spontaneously broken.
In addition, a simultaneous solution of $m_{s}$ and $m_{ps}$ arises 
when the integral in (\ref{s4e15}) and (\ref{s4e16}) vanishes. 
In the Euclidean four-momentum cutoff regularization scheme, this happens at 
$\lvert k\rvert =  m_{s/ps} 
\approx 0.712 \Lambda$. 
The mass spectrum of the scalar and pseudoscalar modes in the NJL model 
thus not only contains the Nambu-Goldstone mode and its chiral partner,
but also an additional finite mass solution, which describes a 
scalar/pseudoscalar mode degeneracy. 

\subsection{$\Gamma_{aPT_2}=\gamma_5$}
\label{s4b}
When the NJL model is modified by the pseudoscalar term $\Gamma_{aPT_2}=\gamma_5$
the resulting extended model is neither Hermitian, nor $\cPT$-symmetric, nor 
chirally symmetric. 
Nevertheless, the model gives rise to real fermion masses as was shown in 
Sec.~\ref{s3d}.
Looking now at the self-consistent mass equation  (\ref{s4e8}) for the 
mesonic bound states, 
we find that evaluating the spinor traces in (\ref{s4e9}) to (\ref{s4e10-2}) for the fermion propagator associated with this model and given in (\ref{s3e30}),
yields
\begin{align}
\label{s4e18}
R^{s}_{aPT_2}(k) &=\int \!\! d^4p \frac{4 (p^2 + p_\mu k^\mu +m^2+g^2)
 }{ ((p+k)^2-m^2+g^2)(p^2-m^2+g^2)},\\
\label{s4e19}
R^{ps}_{aPT_2}(k) &=\int \!\! d^4p \frac{4 (p^2 + p_\mu k^\mu -m^2-g^2)
 }{ ((p+k)^2-m^2+g^2)(p^2-m^2+g^2)},\\
\label{s4e20}
\frac{I_{aPT_2}}{m}
&= \int \!\! d^4 p \, \frac{4(p^2 + p_\mu k^\mu -m^2+g^2) +2 k^2 
 }{ ((p+k)^2-m^2+g^2)(p^2-m^2+g^2)}.
\end{align}
The condition (\ref{s4e8}) for the scalar and pseudoscalar meson modes thus takes 
the form
\begin{align}
\label{s4e21}
0=&\, (k^2-4m^2) \int \!\! d^4p \frac{1}{ ((p+k)^2-m^2+g^2)(p^2-m^2+g^2)} ,\\
\label{s4e22}
0=&\, (k^2+4g^2) \int \!\! d^4p \frac{1}{ ((p+k)^2-m^2+g^2)(p^2-m^2+g^2)},
\end{align}
respectively at 
$k^2 = m_{s/ps}^2$. 
This closely resembles the conditions (\ref{s4e15}) and 
(\ref{s4e16}) of the standard NJL model.

We note, that (unsurprisingly) the appearance of a Nambu-Goldstone mode in the standard 
NJL model breaks down with the inclusion of the chiral symmetry breaking 
term $\Gamma_{aPT_2}$. In its place we find the apparent solution 
to (\ref{s4e22}) that $m_{ps}^2=-4g^2$, i.e. the pseudoscalar solution is a \emph{tachyonic} state with mass $m_{ps}=\pm 2ig$.  
The apparent solution to (\ref{s4e21}) for the scalar meson, however,
remains structurally identical to that of the standard NJL model, cf. (\ref{s4e15}); 
but the effective fermion mass $m$ is now a function of the coupling 
constant $g$ in this modified model, see (\ref{s3e34}). Thus 
$m_{s}^2 = 4m^2 = 4(m_{NJL}^2+g^2)$.
The dynamical generation of fermion mass thus also affects the 
scalar meson.
In addition, the dependence of the fermion mass on the coupling 
constant $g$ also affects the appearance of a simultaneous solution of 
$m_{s}$ and $m_{ps}$ when the momentum integral in (\ref{s4e21}) and 
(\ref{s4e22}) vanishes. 
In fact, it counteracts the explicit $g$ dependence of the momentum integrals, 
leaving the same expression 
that is found in the standard NJL model, cf. 
(\ref{s4e15}) and (\ref{s4e16}). 
We thus find a scalar/pseudoscalar mode degeneracy with the same mass, 
$m_{s/ps} \approx \pm 0.712 \Lambda$ in the Euclidean four-momentum 
cutoff regularization, which is unaffected by the explicit breaking
of chiral symmetry in this model.

Altogether, the bilinear based on $\Gamma_{aPT_2}$, while causing a dynamical 
mass gain of the fermion and the scalar meson, seems to act as a tachyonic 
instability for the pseudoscalar meson. 
The scalar/pseudoscalar mode degeneracy of the standard NJL model remains
unchanged by this extension.
We point out that, while the Nambu-Goldstone mode of the standard NJL model
becomes tachyonic in this extended model, the combination of the non-degenerate
scalar and pseudoscalar meson masses $m_s^2+m_{ps}^2=4m_{NJL}^2$ remains valid.

\subsection{$\Gamma_{aPT_1}=iB_\mu \gamma^\mu$ }
\label{s4c}

Having analyzed the modified NJL model, 
which breaks chiral symmetry explicitly, we now turn to the remaining two 
models in which chiral symmetry is conserved: 
the  non-Hermitian models based on the vector background field 
$\Gamma_{aPT_1}=iB_\mu \gamma^\mu$, which we discuss in this section, and the pseudovector background field
$\Gamma_{PT_1}=i\gamma_5 B_\mu \gamma^\mu$, which is discussed in Sec.~\ref{s4d}. 
As per the results of Sec.~\ref{s3c}, real fermion masses are dynamically generated for all values of the coupling constant in the former, non-$\cPT$-symmetric case. 
In the latter, $\cPT$-symmetric case this happens 
in the finite region of coupling
values smaller than $\tilde{g}_{dyn}\approx 1.183$,
cf. Sec.~\ref{s3a}.

For the NJL model that is extended by an anti-$\cPT$-symmetric scalar background field term $\Gamma_{aPT_1}=iB_\mu \gamma^\mu$ calculating the spinor traces in the terms (\ref{s4e9}) and (\ref{s4e10}), with the fermion
propagator as given in (\ref{s3e23}),
results in the expressions 

\begin{figure*}
\centering
\null\hfill
\subfloat[]{
\includegraphics[width=0.48\textwidth]
{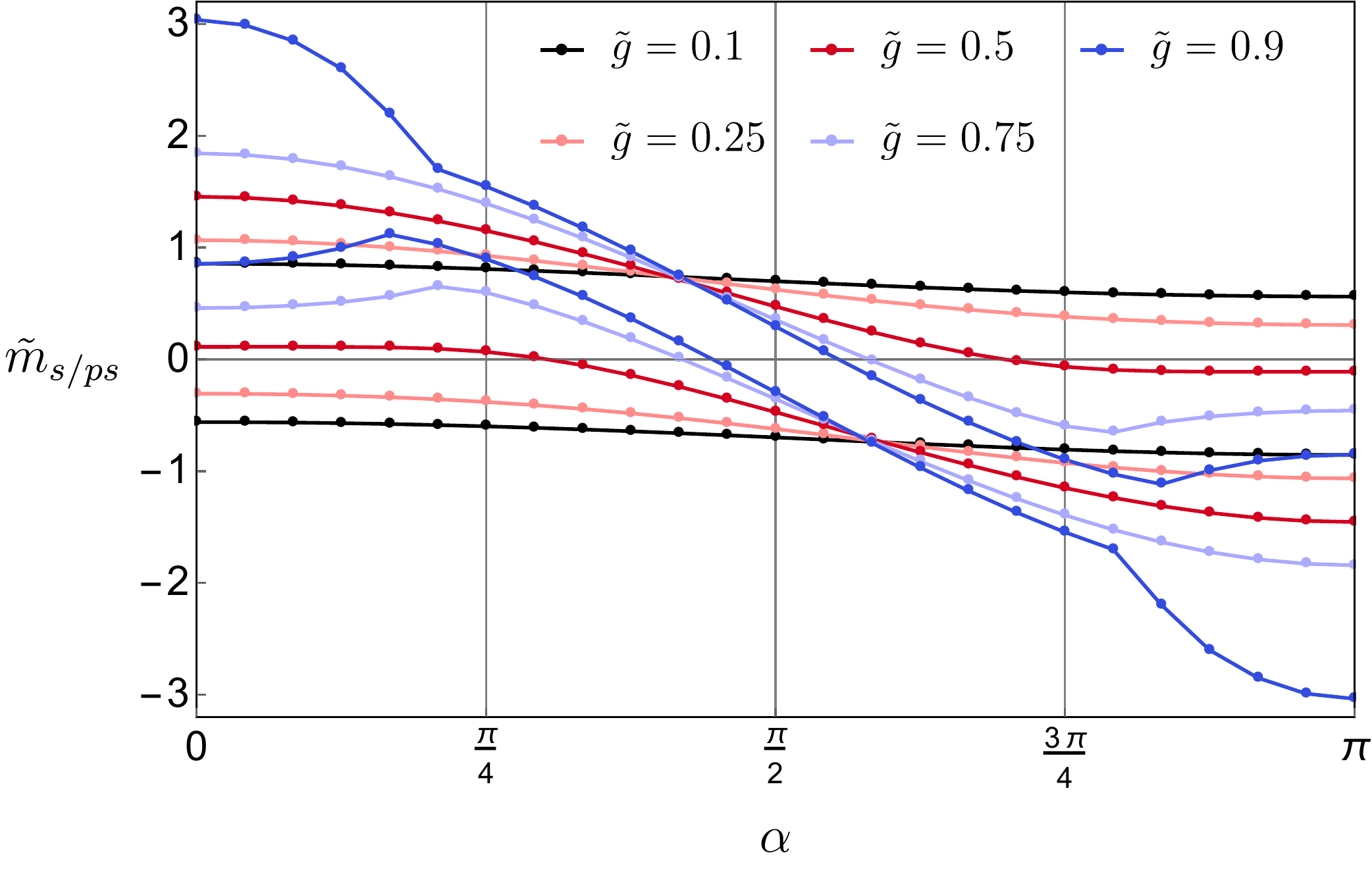}
\label{f8}
}
\hfill
\subfloat[]{
\includegraphics[width=0.48\textwidth]
{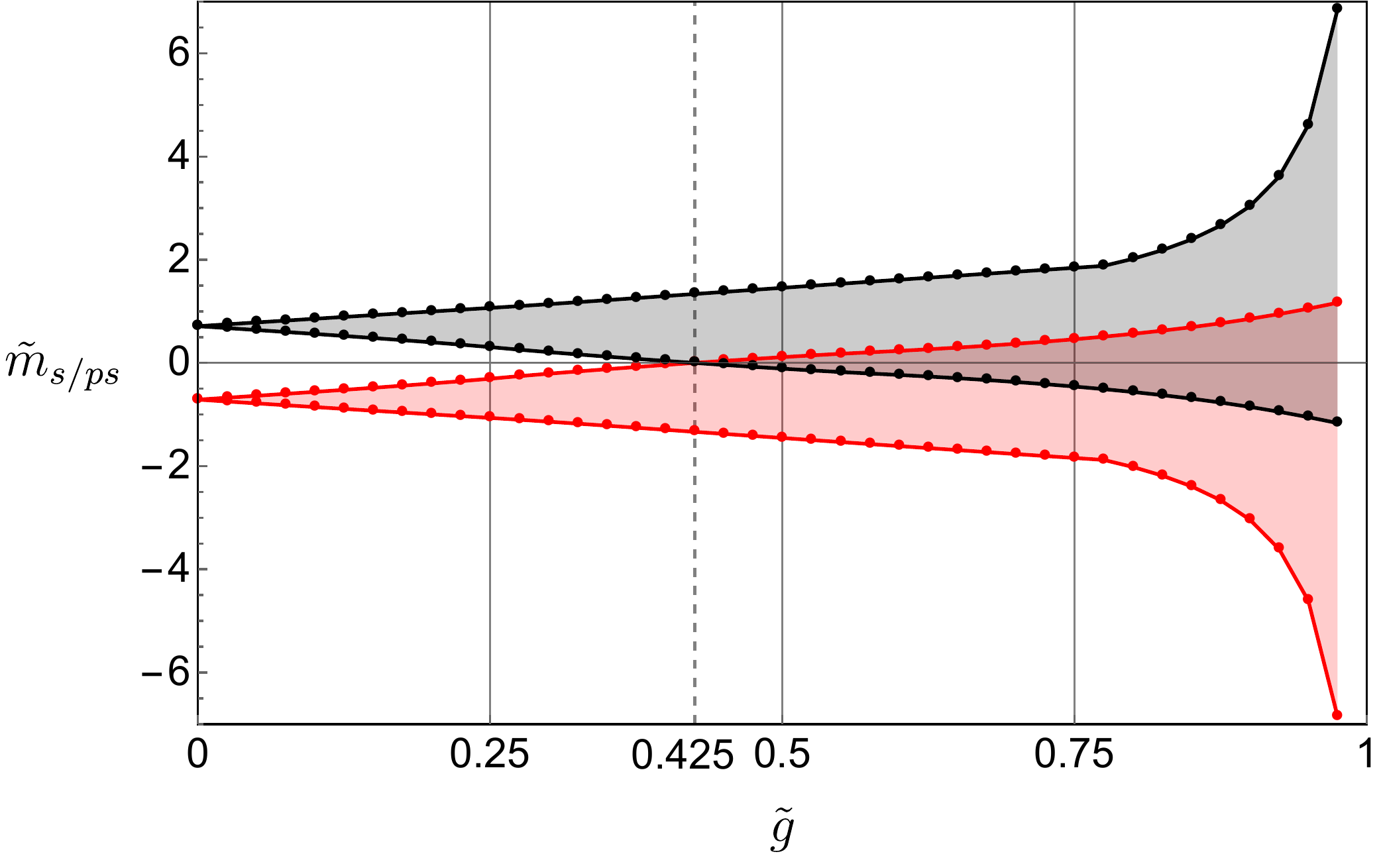}
\label{f9}
}
\caption{ 
In (a) the mass of the scalar/pseudoscalar mode degeneracy in the modified NJL model based on $\Gamma_{aPT_1}$ is shown as function of the angle $\alpha$
between the Euclidean background field and the Euclidean meson momentum for 
fixed values of the coupling constant $\tilde{g}$.
The region of accessible mass over this angle interval is visualized as 
function of $\tilde{g}$ in (b). 
}
\end{figure*}

\begin{widetext}
$$
R^{s}_{aPT_1}(k) =\int \!\! d^4p \frac{
4 (p^2 + p_\mu k^\mu +m^2-g^2B^2 -2ig B_\mu p ^\mu -ig B_\mu k ^\mu)
}{ ((p+k)^2-m^2-g^2B^2 -2ig B_\mu (p+k)^\mu)(p^2-m^2-g^2B^2-2ig B_\mu p^\mu)}
$$
and 
$$
R^{ps}_{aPT_1}(k) =\int \!\! d^4p \frac{
4 (p^2 + p_\mu k^\mu -m^2-g^2B^2 -2ig B_\mu p ^\mu -ig B_\mu k ^\mu)
 }{ ((p+k)^2-m^2-g^2B^2 -2ig B_\mu (p+k)^\mu)(p^2-m^2-g^2B^2
 -2ig B_\mu p^\mu)}.
$$
With the integral $I_{aPT_1}$ in the form (\ref{s4e10-2}) being
$$
\frac{I_{aPT_1}}{m}
= \int \!\! d^4p \frac{
4 (p^2 + p_\mu k^\mu -m^2-g^2B^2 -2ig B_\mu p ^\mu -ig B_\mu k ^\mu) +2k^2
 }{ ((p+k)^2-m^2-g^2B^2 -2ig B_\mu (p+k)^\mu)(p^2-m^2-g^2B^2
 -2ig B_\mu p^\mu)},
$$
we obtain the following form for the self-consistent mass equation 
(\ref{s4e8}) for the scalar and 
pseudoscalar modes:
\begin{align}
\label{s4e27}
0=&\, (k^2-4m^2) \int \!\! d^4p \frac{1}{ ((p+k)^2-m^2-g^2B^2 
-2ig B_\mu (p+k)^\mu)(p^2-m^2-g^2B^2 -2ig B_\mu p^\mu)} ,\\
\label{s4e28}
0=&\, k^2 \int \!\! d^4p \frac{1}{ ((p+k)^2-m^2-g^2B^2 -2ig B_\mu (p+k)^\mu)
(p^2-m^2-g^2B^2 -2ig B_\mu p^\mu)}.
\end{align}

Both of the apparent solutions that $m_{ps}^2=0$ and $m_{s}^2 = 4m^2$ are structurally identical to those of the standard NJL model, cf. Eqs. 
(\ref{s4e15}) and (\ref{s4e16}).
The existence of a massless Nambu-Goldstone mode remains unaffected throughout 
the extension of the model by the non-Hermitian anti-$\cPT$-symmetric term $\Gamma_{aPT_1}$. 
One can understand this, as chiral 
\pagebreak
\end{widetext}
symmetry continues to be broken spontaneously.

The mass of the scalar meson, $m_{s} = 2 m$, remains structurally
identical as well, 
although once again the fermion  mass $m$ now depends on the coupling constant $g$, cf. Sec.~\ref{s3c}. 
The dynamical mass generation of the fermion is thus reflected in the mass of the scalar meson.

Furthermore, the appearance of the identical four-momentum integrals in both 
conditions (\ref{s4e27}) and (\ref{s4e28}) suggests that an additional 
simultaneous solution of $m_{ps}$ and $m_{s}$ can be found here as well.
However, the appearance of a term $B_\mu k^\mu$ in the denominator of this 
integral indicates that this solution depends on the angle between 
the background field $B_\mu$ and the meson momentum $k_\mu$.
The roots of the momentum integral can be determined numerically in the 
Euclidean four-momentum cutoff regularization scheme for various values of the 
angle $\alpha$ between the Euclidean background field $B_E$ 
(where $B_0= iB_4$) and the Euclidean meson momentum $k_E$ 
(where $k_0= ik_4$).
Figure~\ref{f8} shows the behavior of the degenerate meson mass $\tilde{m}_{s/ps}$
as a function of the angle $\alpha$ for small values of the scaled coupling 
constant $\tilde{g}= g\lvert B_E\rvert \Lambda^{-1}<1$, which describes the 
amplitude of the vector background field.
In Fig.~\ref{f9} the extremal values of the meson mass in the angular 
interval are visualized as a function of the coupling $\tilde{g}$
in the same region. 
Since the meson mass is a continuous function in $\alpha$, all mass values 
between these extrema can be found for some value of $\alpha$. 
This is shown in Fig.~\ref{f9} as shaded regions.
We note that for coupling values larger than 
$\tilde{g}\approx 0.425$
arbitrarily small (or vanishing) meson masses $m_{\/ps}$ can be realized.

Overall, we find a Nambu-Goldstone mode and a scalar meson with mass 
$m_{s} =  2m$ reflecting the dynamical mass generation of the 
fermion, plus the existence of an additional scalar/pseudoscalar degeneracy in the modified NJL model based on 
$\Gamma_{aPT_1}=iB_\mu \gamma^\mu$. 
The simultaneous solution shows an intricate dependence on the amplitude
of the scalar background field and its angle relative to the meson momentum. 
For sufficiently large values of the 
coupling this mass solution may be arbitrarily small (or vanish entirely),
but such values of the coupling also generate a significant fermion mass in the model.

\subsection{$\Gamma_{PT_1}=i\gamma_5 B_\mu \gamma^\mu$}
\label{s4d} 

\begin{figure*}
\centering
\null\hfill
\subfloat[]{
\includegraphics[width=0.48\textwidth]
{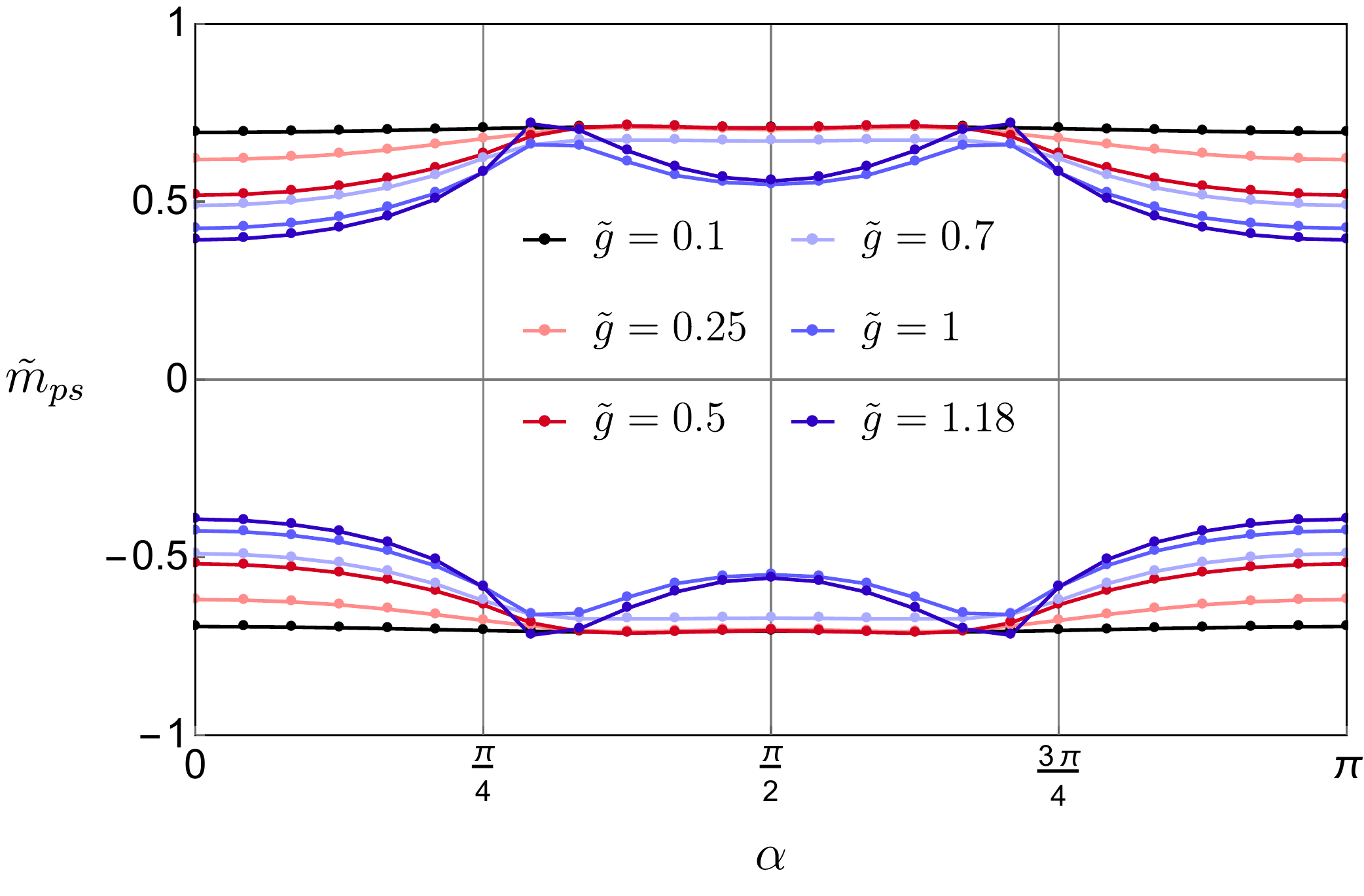}
\label{f10}
}
\hfill
\subfloat[]{
\includegraphics[width=0.48\textwidth]
{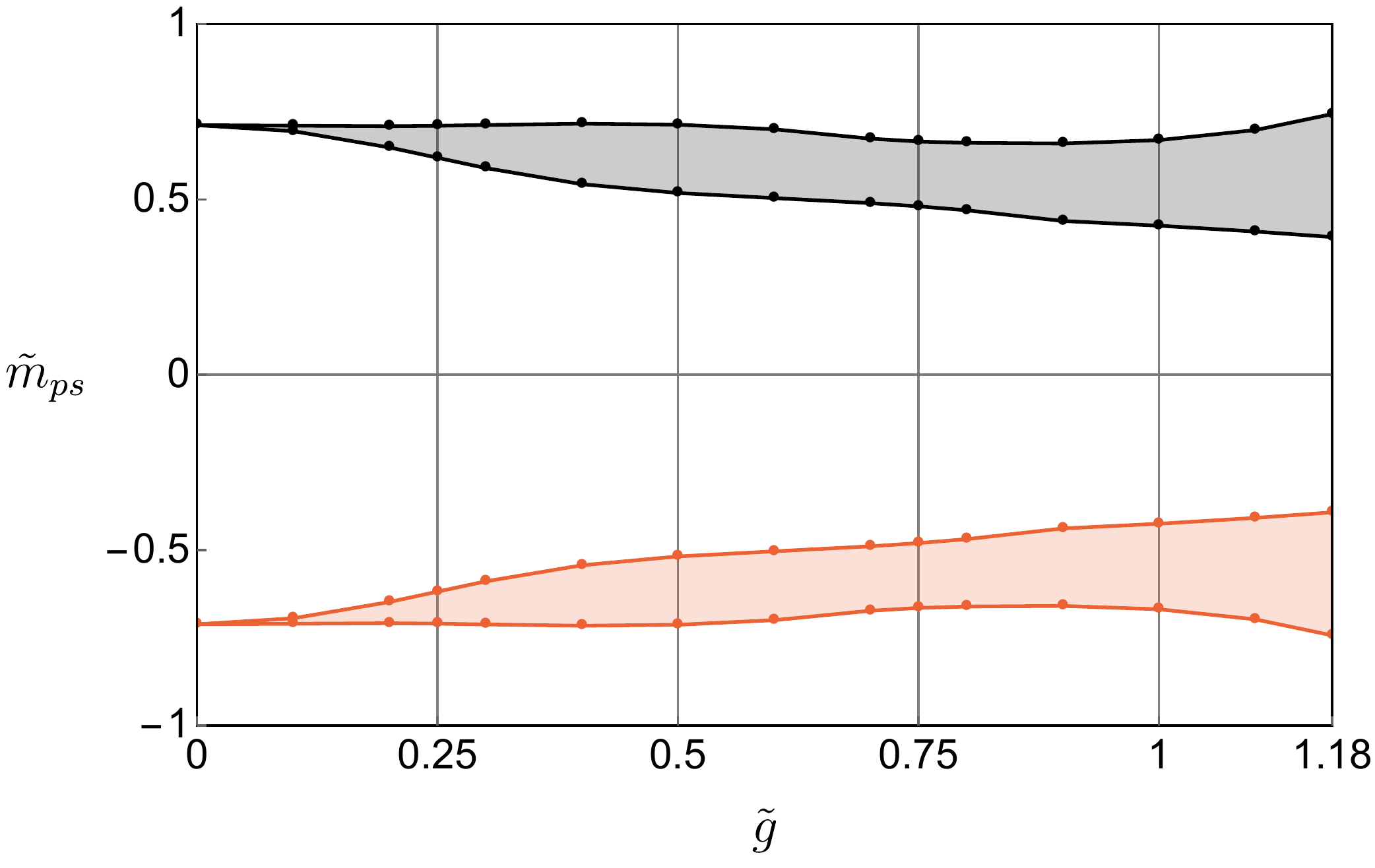}
\label{f11}
}
\caption{ 
In (a) the mass of the pseudoscalar meson in the modified NJL model based on $\Gamma_{PT_1}$ is shown as function of the angle $\alpha$
between the Euclidean background field and the Euclidean meson momentum for 
fixed values of the coupling constant $\tilde{g}$.
The region of accessible mass over this angle interval is visualized as 
function of $\tilde{g}$ in (b). }
\end{figure*}

When the standard NJL model is extended by a $\cPT$-symmetric pseudovector background field
$\Gamma_{PT_1}=i\gamma_5 B_\mu \gamma^\mu$, its rationalized fermion propagator is given by Eq.~(\ref{s3e3-2}),
which does not simplify as much as in the models discussed previously.
The evaluation of the spinor traces in the terms (\ref{s4e9}), (\ref{s4e10}) and (\ref{s4e10-2}) thus yields somewhat cumbersome results.
However, when inserted into the self-consistent meson mass equation (\ref{s4e8}) some contributions simplify and the conditions for the scalar and pseudoscalar meson masses become
\begin{widetext}
\begin{align}
\label{s4e30}
0= \int \!\! d^4p \,
\frac{ (k^2-4m^2) A(p,k) + 8g^2 B(p,k)}
{\bigl[((p+k)^2 - m^2 - g^2 B^2)^2 - 4 g^2 m^2 B^2
+ 4 g^2 (B_\mu (p+k)^\mu)^2\bigr]
\bigl[(p^2 - m^2 - g^2 B^2)^2 - 4 g^2 m^2 B^2
+ 4 g^2 (B_\mu p^\mu)^2\bigr]}
\end{align}
and
\begin{align}
\label{s4e31}
0= \int \!\! d^4p \,
\frac{ k^2 A(p,k) + 8g^2 B(p,k)}
{\bigl[((p+k)^2 - m^2 - g^2 B^2)^2 - 4 g^2 m^2 B^2
+ 4 g^2 (B_\mu (p+k)^\mu)^2\bigr]
\bigl[(p^2 - m^2 - g^2 B^2)^2 - 4 g^2 m^2 B^2
+ 4 g^2 (B_\mu p^\mu)^2\bigr]}
\end{align}
respectively at $k^2 = m_{s/ps}^2$, where
\begin{align}
\label{s4e32}
A(p,k)=&
((p+k)^2-m^2+g^2B^2) (p^2-m^2+g^2B^2)
+4g^2 [B^2 p_\mu (p+k)^\mu - (B_\mu p^\mu) (B_\nu (p+k)^\nu)],\\
\label{s4e34}
B(p,k)=& 
2(p_\mu k^\mu)(B_\nu k^\nu)(B_\alpha p^\alpha)
-B^2 (p_\mu k^\mu)^2
+p^2 k^2 B^2
-k^2 (B_\mu p^\mu)^2
-p^2 (B_\mu k^\mu)^2 .
\end{align}
\end{widetext}

When considering the vector products in $B(p,k)$, we notice an overall 
proportionality to $\lvert k \rvert^2$.  
Thus the condition (\ref{s4e31}) for the pseudoscalar meson mass has an apparent 
solution that $m_{ps}^2 = 0$. 
Once again, a Nambu-Goldstone mode appears as a manifestation of chiral symmetry in this modified NJL model.
A second solution for the pseudoscalar meson mass $m_{ps}$ can be found by evaluating the 
momentum integral in  (\ref{s4e31}) numerically in the Euclidean 
four-momentum cutoff regularization scheme. 
Similar to the case of the model based on the vector background field 
$\Gamma_{aPT_1}$ discussed in Sec.~\ref{s4c}, this 
solution depends on the amplitude of the (pseudovector) background field
$B_\mu$ and its angle with the pion momentum $k_\mu$.
Figure \ref{f10} shows the behavior of this pseudoscalar meson mass as a function of 
the angle $\alpha$ between the Euclidean background field $B_E$ 
(where $B_0= iB_4$) and the Euclidean meson momentum $k_E$ 
(where $k_0= ik_4$) at fixed values of the coupling constant 
$\tilde{g}=g\lvert B_E \rvert \Lambda^{-1}$ for which a fermion mass is 
dynamically generated, i.e. $\tilde{g}<\tilde{g}_{dyn}\approx 1.183$.
In Fig.~\ref{f11} the extremal values of the pseudoscalar meson mass in the angular 
interval are visualized as a function of the coupling constant $\tilde{g}$
in the same region. 
Since again the mass is a continuous function of $\alpha$, all values 
between these extrema can be obtained for some value of $\alpha$, which is
indicated in Fig.~\ref{f11} as shaded regions.
Notice that independent of the values of $\alpha$ and $\tilde{g}$ the 
second pseudoscalar mass solution $m_{ps}$ has finite real values.

\begin{figure*}
\centering
\null\hfill
\subfloat[]{
\includegraphics[width=0.48\textwidth]
{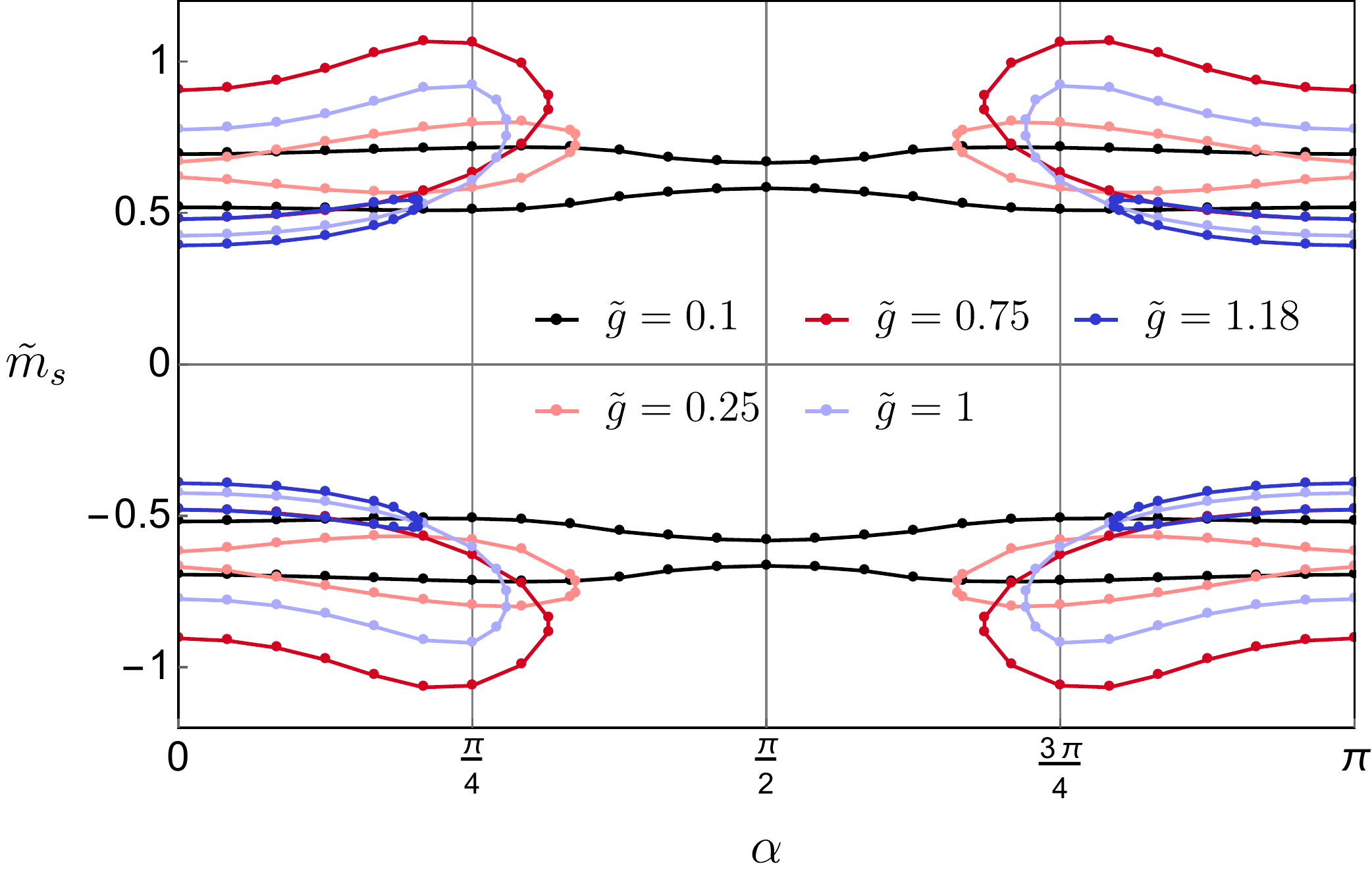}
\label{f12}
}
\hfill
\subfloat[]{
\includegraphics[width=0.48\textwidth]
{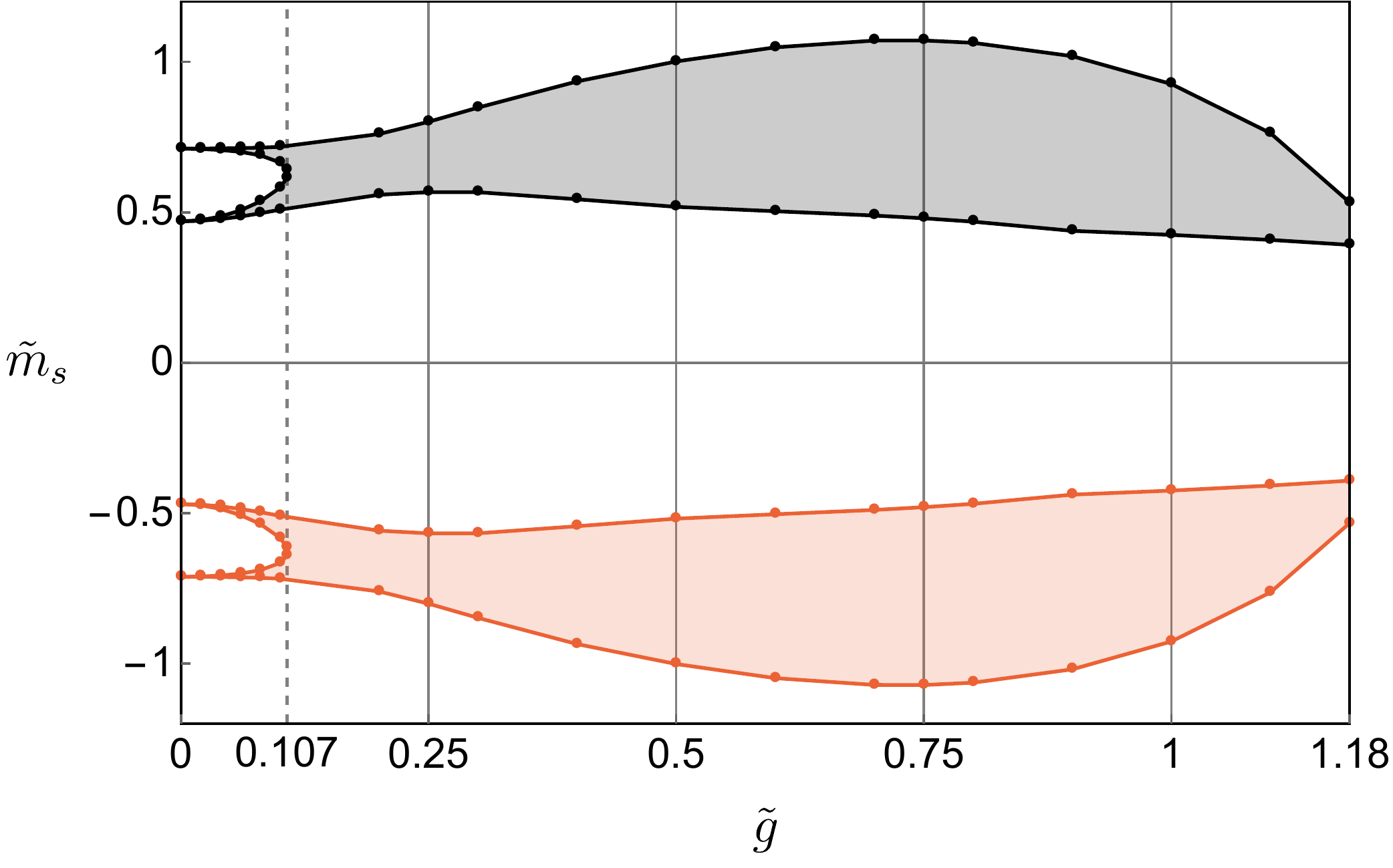}
\label{f13}
}
\caption{ 
In (a) the masses of the scalar meson in the modified NJL model based on $\Gamma_{PT_1}$ are shown as function of the angle $\alpha$
between the Euclidean background field and the Euclidean meson momentum for 
fixed values of the coupling constant $\tilde{g}$.
The region of accessible mass over this angle interval is visualized as 
function of $\tilde{g}$ in (b).}
\end{figure*}
For the scalar meson mass as determined by condition (\ref{s4e30}), 
we do not generally find an apparent solution that factors out of the momentum 
integral. 
In particular, we do not find a scalar meson mass that 
is only proportional to the fermion mass $m$ of the model, in contrast to
the solutions of the other non-Hermitian models discussed so far.
We also do not generally find a simultaneous solution of the scalar
and pseudoscalar meson modes.
The two solutions $m_s$ that one can find by evaluating 
(\ref{s4e30})  numerically  in the Euclidean four-momentum cutoff regularization scheme \emph{both} 
depend on the angle $\alpha$ between  $B_E$ and $k_E$, and on the value of the scaled coupling constant 
$\tilde{g}=g\lvert B_E \rvert \Lambda^{-1}$, which is proportional to the 
amplitude of the background field.
Their behavior is shown in Fig.~\ref{f12} as a function of the angle $\alpha$
and for fixed values of $\tilde{g}$.
We point out that only for small coupling values below
$\tilde{g}\approx 0.107$ 
do real mass solutions $m_{s}$ exist for all angles $\alpha$. 
For larger coupling values the solutions around $\alpha=\pi/2$ begin to 
break down.
However, the region of real masses $m_s$ never breaks down 
entirely and real-valued solutions can always be found for some value of 
$\alpha$.

The mass of the scalar mode is a continuous function of 
$\alpha$ and any real mass value between the extremal values over the angle interval is realized at some value of $\alpha$.
The behavior of the extremal values is visualized in Fig.~\ref{f13} as a function of coupling constant values for which a fermion mass is dynamically 
generated ($\tilde{g}<\tilde{g}_{dyn}\approx 1.183$). The 
region of accessible mass values in between minima and maxima is indicated by the shaded area.
As remarked above, the two scalar meson mass solutions remain real
and separate solutions at all angles $\alpha$ for coupling values below 
$\tilde{g}\approx 0.107$. For higher coupling values the solutions combine 
as shown in Fig.~\ref{f12}.

\emph{Special case $\alpha=0$}.
We remark that when the pseudovector background field is parallel 
to the meson momenta, i.e. $\alpha=0$, the contribution $B(p,k)$ in the 
conditions (\ref{s4e30}) and (\ref{s4e31}) vanishes.
In this special case we once again find the apparent solutions 
$m_{ps}^2=0$ and $m_{s}^2=4m^2$ as in the case of the standard NJL model 
and the extended model based on the vector field $\Gamma_{aPT_1}$.

We further note that $\alpha=0$ is a special case of the underlying 
modified Dirac theory as well:
The dispersion relation of the model with Hamiltonian density
$$
\cH=\bar\psi (-i \gamma^k \partial_k +ig \gamma_5 B_{\mu}\gamma^{\mu} )
\psi
$$
has the form 
$$
0 = (p^2-g^2B^2)^2 +4g^2(B_\mu p^\mu)^2
$$
at $p^2=m^2$.
Thus the effective mass of the modified Dirac fermion has the form 
$m^2= g^2B^2 \exp[\mp 2i\alpha]$.
This generally describes complex mass solutions, as noted in \cite{bkb},
but in the special case of $\alpha=0$ we find a purely real fermion mass of $m^2= g^2B^2$.

Overall, we find a Nambu-Goldstone mode for the pseudoscalar meson of this chirally 
symmetric theory as well as an additional finite pseudoscalar meson mass solution that 
depends in particular on the angle $\alpha$ between its momentum and the pseudovector background field.
Independently, we have found two scalar meson mass solutions
which display an intricate dependence on the angle $\alpha$ as well and which,
in particular, may break down depending on the specific value of $\alpha$
and the coupling constant $g$.

\section{Concluding remarks}
\label{s5}

It is surprising that the free Dirac equation in 3+1 dimensions, augmented by any $\cPT$-symmetric bilinears  does not lead to a real energy spectrum for the fermions 
unless a bare quark mass is present  \cite{bkb}. (This is counterintuitive, since for  bosonic systems, a region in which $\cPT$-symmetry is unbroken  leading to a real mass spectrum 
can usually be found \cite{b}.) However, it was shown in \cite{fbk} that in a fermionic system such as the NJL model, which contains additional two-body interactions 
a phase of unbroken $\cPT$ symmetry having real energies can occur within a specific parameter range. In that case, the $\cPT$-symmetric bilinear included 
into the Hamiltonian density $\cH$ given in (\ref{s2e2}) was $\Gamma_{PT_1}$. Having determined that such a Hamiltonian can lead to an additional 
dynamical mass generation, mimicking the effect of having a bare fermion mass $m_0$, we have examined in detail in this paper the effects of including all other possible 
non-Hermitian bilinear interactions into the Hamiltonian, not only in view of their ability to generate a real fermion spectrum, but also in view of the implications for constructing 
composite particles. In particular, our interest has been to understand the roles of  both $\cPT$ and chiral symmetry, which are both relevant in this context. We have thus considered
non-Hermitian scalar ($\Gamma_{aPT_3}$), pseudoscalar ($\Gamma_{aPT_2}$), vector ($\Gamma_{aPT_1}$),  pseudovector ($\Gamma_{PT_1}$) and tensor ($\Gamma_{PT_2}$) 
additions to the NJL Hamiltonian. 

We find the following results:
 
(a) $\cPT$ symmetry is not necessary for real fermion mass solutions to exist, rather 
the two-body interactions supersede the non-Hermitian bilinear effects. To be specific, real solutions for the fermion mass occur for all interactions, with the exception of 
$\Gamma_{aPT_3}$. While the presence of $\Gamma_{PT_1}, \Gamma_{aPT_1}$ and $\Gamma_{aPT_2}$ (pseudovector, vector and pseudoscalar contributions) can lead to an increase in the dynamically generated mass, mimicking the role of a current quark mass, this is not true when $\Gamma_{PT_2}$ is included: then a loss in mass results.

(b) The role of chiral symmetry becomes most evident in the mesonic sector. Both chirally symmetric bilinears,  $\Gamma_{aPT_1}$ (vector) and $\Gamma_{PT_1}$ (pseudovector), 
as mentioned already in (a), allow for the dynamical generation of fermion masses. In addition, the Goldstone mode remains unaffected, lying again at zero, manifesting the chiral 
symmetry of the model.  For the vector interaction, the scalar  (or $\sigma$) meson gains mass dynamically as does the fermion, while a second degenerate solution for the composite 
scalar and pseudoscalar modes depends on the relation of the background field and the meson momentum. It is possible to generate small masses at values of the coupling $\tilde{g}
>0.425$, but this again would imply that there is a  significant fermion mass. 

For the pseudovector interaction no general scalar/pseudoscalar degeneracy is found. Furthermore, the additional pseudoscalar and scalar meson masses depend on the
relation of the background field and the meson momentum and remain "large" (of the same size as the degenerate state of the standard NJL model)

The inclusion of a pseudoscalar bilinear leads to a tachionic  pseudoscalar mode (which is unphysical) while the scalar mode
gains mass dynamically in the same fashion as the fermion, the degenerate solution remaining unaffected.

We thus conclude that the mechanism of mass generation due to four-fermion interactions can occur in many different ways. There seems to be no argument for discarding 
non-Hermitian bilinears that also give rise to the generation of real fermion masses. With the exception of $\Gamma_{aPT_3}$, all other forms do this when bare fermion masses 
are absent. However, in order to give the Goldstone mode a finite mass, another mechanism is required. An additional scalar/pseudoscalar degenerate mode can often be found 
that is tunable with the coupling strength. 

We point out that this analysis is applicable generally to Hamiltonians that include four-fermion interactions, as used for example in discussions of the generation of
neutrino masses \cite{ams,ms}.

\end{document}